\documentclass[onecolumn]{aa}
\usepackage{graphicx,natbib}

\begin{document}

\title{Extending the Shakura-Sunyaev approach
to a strongly magnetized accretion disc model}
 
\author 
{V.I. Pariev\inst{1}\inst{2}
       \and
       E.G. Blackman\inst{1} 
       \and 
       S.A. Boldyrev\inst{3}}
	
       \offprints{V.I. Pariev}
	
       \institute{		 	
Department of Physics and Astronomy, University of Rochester, 
Rochester, NY 14627, USA 
    \and
P.N. Lebedev Physical Institute, Leninsky Prospect 53, Moscow
117924, Russia 
     \and		
Department of Astronomy and Astrophysics, University of Chicago, 
5640 South Ellis Avenue, Chicago, IL 60637, USA
}

\date{Received; accepted}

\titlerunning{A strongly magnetized accretion disc model}
\authorrunning{V.I. Pariev, et al.}

\abstract{ 
We develop a model of thin turbulent accretion discs supported
by magnetic pressure of turbulent magnetic fields.
This applies when the turbulent kinetic and magnetic energy
densities are greater than the thermal energy density
in the disc. Whether such discs survive in nature or not
remains to be determined,  but here we simply 
demonstrate that self-consistent solutions
exist when the $\alpha$-prescription for the viscous stress, 
similar to that of the original 
Shakura--Sunyaev model, is used. 
We show that $\alpha \sim 1$ for the strongly 
magnetized case and  we calculate the radial structure
and emission spectra from the disc in the regime when it is optically
thick.
Strongly magnetized optically thick discs can
apply to the full range of disc radii for objects  $\la 10^{-2}$ of the
Eddington luminosity or for the outer parts of discs in higher luminosity
sources. 
In the limit that the magnetic pressure is equal to the thermal 
or radiation pressure, our strongly magnetized disc 
model transforms into the Shakura--Sunyaev model with $\alpha=1$.
Our model produces spectra quite similar to those of standard 
Shakura--Sunyaev models. In our comparative study, we also 
discovered a small discrepancy in the spectral calculations 
of Shakura \& Sunyaev (1973). 

}

\maketitle

\newcommand{\bvec}[1]{\mbox{\boldmath$#1$}}

\section{Introduction}\label{sec1} 

The well known and most widely used model of the accretion disc was proposed
and elaborated by \citet{shakura72} and 
\citet{shakura73}. In this model the disc is vertically supported 
by the thermal pressure \citep{shakura73}.
Turbulent viscosity is invoked in the Shakura--Sunyaev model 
to explain the angular momentum transfer required by the 
accretion flow. As originally pointed out in \citet{lyndenbell69} and  
\citet{shakura73} a magnetic field can also contribute to the angular momentum 
transport. A robust mechanism of the excitation of magnetohydrodynamical (MHD) 
turbulence was shown to operate in accretion discs due to the magneto-rotational
(MRI) instability \citep{balbus98}. The growth of the MRI leads to the excitation 
of turbulent magnetic fields and self-sustained MHD turbulence. The contribution 
of Maxwell stresses to the transport of angular momentum is usually larger
than Reynolds stresses. However, the magnetic energy observed in many numerical 
experiments was smaller than the thermal energy of the gas in the disc 
\citep{brandenburg98}. Simulations of the non-linear stage of MRI 
are typically local simulations in a shearing box 
of an initially uniform small part of the disc. 
Attempts to expand the computational domain to include a wider area of radii and 
azimuthal angle \citep*{hawley01a, hawley01b, armitage01} are underway.
However, even before the recent focus on the MRI \citet*{shibata90}  
observed the formation of transient
low $\beta$ state in a shearing box simulations 
of the non-linear Parker instability in an accretion disc.

Vertical stratification has 
been considered in the shearing box approximation \citep{brandenburg95,
miller00}. In particular, \citet{miller00} investigated discs with initial 
Gaussian density profiles supported by thermal pressure. 
The initial seed magnetic field grows and starts to contribute 
to the vertical pressure balance. The computational domain extends 
over enough vertical scale heights to enable \citet{miller00} to simulate the 
development of a magnetically dominated corona above the disc surface.
In the case of an initial axial magnetic field, \citet{miller00} observed that 
the saturated magnetic pressure dominates thermal pressure not only in the 
corona but everywhere 
in the disc. As a consequence, the thickness of the disc increases until it 
reaches the axial boundaries of the computational box. 
The formation of low 
$\beta$ filaments in magnetized tori was also observed in global MHD simulations
by \citet*{machida00}.
Although further 
global MHD simulations of vertically stratified accretion discs are needed, 
this numerical evidence suggests that magnetically dominated thin discs 
may exist.

Previously, analytic
models of thin accretion discs with angular momentum transfer due to 
magnetic stresses were considered by \citet{eardley75} and \citet{field93a,
field93b}. Both these works included magnetic loops 
with size of the order of the disc thickness. In \citet{eardley75}, the 
magnetic loops were confined to the 
disc. Loop stretching by differential rotation was balanced 
by reconnection. The reconnection speed was a fraction 
of the Alfv\'en speed. Radial magnetic flux was considered as a 
free function of the radius. Vertical equilibrium and heat transfer were 
treated as in \cite{shakura73}, with the addition of the magnetic 
pressure in the vertical support. No self-consistent magnetically dominated 
solutions were found in model of \citet{eardley75}. 

In contrast, dominance of the 
magnetic pressure over the thermal and radiation pressure was postulated 
from the beginning by \citet{field93a,field93b} and verified at the
end of their work. These authors assumed that the ordered 
magnetic field in the disc, amplified by differential 
rotation, emerges as loops above the surface of the disc due to Parker 
instability. Because the radial magnetic field in the disc has
an intially sectorial structure, the loops above the disc  come to close 
contact and reconnect. All dissipation of magnetic field occurs in the
corona in the model of \citet{field93a,field93b}. Such a corona was assumed
to be consisting of electrons and some fraction of positrons and no
outflow from the disc is present.
Electrons and positrons are accelerated to relativistic energies 
at the reconnection sites
in the disc corona and subsequently emit synchrotron and inverse Compton 
photons. Because reconnection was assumed to occur at loop tops,   
\citet{field93a,field93b} found that up to 70 per cent of 
the energy released in reconnection events in the corona will be deposited 
back to the surface of the disc in the form of relativistic particles 
and radiation. Only thin surface of optically thick disc is heated and 
cools by the thermal emission, which is the primary source of soft 
photons for the inverse Compton scattering by relativistic particles in 
the corona. 

Since the characteristic velocity of rise of the loops of 
the buoyant magnetic field is of the order of the Alfv\'en speed, it
takes about the time of a Keplerian revolution for the loop of the 
magnetic field to rise (e.g., \citealt{beloborodov99}). 
This is also about the characteristic dissipation
time of the magnetic field in shocks inside the disc (see section~\ref{sec2}).
The  model we explore here differs from that of \citet{field93a,field93b}
in that the dissipation of the magnetic energy occurs essentially inside 
the disc and the heat produced is transported to the disc surface and 
radiated away. 
Observations of hard X-ray flux indicate the presence
of hot coronae where a significant fraction of the total accretion power is 
dissipated. For example, 
the X-ray band carries a significant fraction of the total luminosity of
Seyfert nuclei: the flux in the 1--10~keV band is about $1/6$ of the 
total flux from {\it IR} to X-rays, and the flux in 1--500~keV band is 
about 30--40 per cent of the total energy output \citep*{mushotzky93}.
Another example is the low/hard state of galactic black hole sources,
where the borad band spectrum is completely dominated by a hard X-ray
power law, rolling over at energies of $\sim 150\,\mbox{keV}$ 
\citep{nowak95, done02}. Also, in the so called very high state,
some of galactic black hole X-ray sources show both thermal and non-thermal
(power law) components, with the ratio of non-thermal to total luminosity
of 20--40 per cent \citep{nowak95}. 
Reconnection events and particle acceleration
should also happen in rarefied strongly magnetized corona of the disc in 
our model and could cause observed X-ray flaring events. 
However, we do not consider the
coronal dynamics here, and instead just focus on the structure and the 
emission spectrum of the disc itself. 

Models of magnetized accretion discs with externally imposed
large scale vertical magnetic field and anomalous magnetic field diffusion 
due to enhanced turbulent diffusion have also been considered 
\citep{shalybkov00, campbell00, ogilvie01}. 
The magnetic field in these models was strong enough 
to be dynamically important. But those models 
are limited to the subsonic turbulence in the disc and the viscosity and 
magnetic diffusivity are due to hydrodynamic turbulence. Angular momentum 
transport in those models are due to the large scale global magnetic fields. 
Both small scale and large scale magnetic fields should be present in real 
accretion discs. Here we consider the possibility 
that the magnetic field has dominant 
small scale component, that is magnetic field inside the disc 
consists mostly of loops with size less than or comparable to the thickness
of the disc.

We consider vertically integrated equations describing the 
radial structure of the magnetically dominated turbulent accretion disc and provide 
the solutions for the
radial dependences of the averaged quantities in section~\ref{sec2}. 
In section~\ref{sec3} we analyse the conditions for a magnetically 
supported disc to be self-consistent. In section~\ref{sec4} we calculate 
thermal emission spectra of magnetically supported disc taking into account 
scattering by free electrons.

\section{Radial Disc Structure}\label{sec2}

Here we neglect effects of general relativity and do not consider the behaviour 
of the material closer than the radius of the 
innermost circular stable orbit $r_{\rm {s}}$. We assume a non-rotating black hole with 
$r_{\rm {s}}=3 r_{\rm {g}}$, where the gravitational radius of the black hole of mass 
$M=M_8 \times 10^8 {\rm M}_{\sun}$ is
$\displaystyle r_{\rm {g}}=2GM/c^2 = 3\times 10^{13}\,\mbox{cm}\, M_8$.
We assume that accretion occurs 
in the form of a geometrically thin accretion disc and verify this assumption 
in section~\ref{sec3}. 
We consider a disc of half-thickness $H$, surface density $\Sigma$, 
averaged over the disc thickness volume density $\rho=\Sigma/(2H)$,  
accretion rate ${\dot M}$, and radial 
inflow velocity $v_r$, $v_r>0$ for the accretion. We take 
$\Omega_{\rm {K}}=(GM/r^3)^{1/2}=7.2\times 10^{-4}\,\mbox{s}^{-1} M_8^{-1} 
(r_{\rm {g}}/r)^{3/2}$
to be the angular Keplerian frequency at the radial distance $r$ from the black 
hole. Then, equation of mass conservation reads
\begin{equation}
{\dot M}=2\pi r \Sigma v_r\label{eqn1}\mbox{.}
\end{equation}
In the stationary state ${\dot M}$ does not depend on time.
Equation of angular momentum conservation is (e.g., \citet{shapiro83})
\begin{equation}
f_{\phi}(r) 2H 2\pi r^2 = {\dot M}\left(\sqrt{GMr}-\zeta\sqrt{GMr_{\rm {s}}}\right)
\label{eqn2}\mbox{,}
\end{equation}
where $f_{\phi}$ is the tangential stress at a radius $r$, which 
acts on the inner part of the disc, factor $\zeta$ accounts for the unknown 
torque acting on the disc at the inner edge $r_{\rm {s}}$. 
In the standard Shakura--Sunyaev
model, $\zeta=1$, which corresponds to a zero torque inner boundary. The 
tangential stress at the innermost stable orbit, $f_{\phi}(r_{\rm {s}})$ is related to
the parameter $\zeta$ as follows
\begin{equation}
f_{\phi}(r_{\rm {s}}) 2H 2\pi r_{\rm {s}}^2=(1-\zeta){\dot M}
\sqrt{GM r_{\rm {s}}}\label{eqn3}\mbox{.}
\end{equation}
The stress $f_{\phi}(r_{\rm {s}})>0$, if the torque spins up the gas at 
$r_{\rm {s}}$, and $f_{\phi}(r_{\rm {s}})<0$, if the torque retards the rotation of the gas. 
Since the magnetic and turbulent pressures exceed the thermal pressure, by 
assumption,
we neglect thermal pressure contribution to the vertical pressure balance. 
Assuming equipartition between turbulent and magnetic pressure 
in supersonic MHD turbulence \citep*{stone98, stone99, miller00} it 
is easy to show that the equation of vertical equilibrium is solved to give 
approximately the result
\begin{equation}
H=\frac{v_{\rm {A}}}{\Omega_{\rm {K}}}=\frac{1}{\Omega_{\rm {K}}}
\sqrt{\frac{HB^2}{2\pi\Sigma}}
\label{eqn4}\mbox{,}
\end{equation}
where $\displaystyle v_{\rm {A}}=\frac{B}{(4\pi\rho)^{1/2}}$ is the average Alfv\'en 
velocity in the disc. 
We use the '$\alpha$--prescription' for the magnetic viscosity in the disc, i.e. 
taking the tangential stress to be proportional to the sum of the magnetic pressure, 
$\displaystyle  \frac{B^2}{8\pi}$, and the turbulent pressure, $\rho v_{\rm {t}}^2/2$. 
We do not consider global transport of angular momentum and energy
across the disc due from large scale magnetic fields. We assume the magnetic 
field to be sufficiently tangled on scales of order of the thickness
of the disc, such that large scale field is small compared to this tangled 
field produced by turbulence.
In equipartition, $\displaystyle \frac{B^2}{8\pi}\approx \rho v_{\rm {t}}^2/2$
and $\alpha$--prescription becomes 
\begin{equation}
f_{\phi}=\alpha \frac{B^2}{4\pi}\label{eqn5}\mbox{.}
\end{equation}

The dissipation of the magnetic and kinetic energy 
causes heat input in the disc which is balanced by 
the heat losses due to radiation. If the cooling is efficient 
enough such that the time of radial advection of the heat due to the accretion flow
is much longer than a few Keplerian periods, the heating and cooling balance, and 
establish an equilibrium kinetic temperature in the disc. However,  
we assume that this temperature is insufficient for the associated thermal pressure to 
contribute significant vertical support. 
The total radiated energy
from the unit surface of the disc will be the same as in the standard Shakura--Sunyaev
model. This energy flux is independent of the viscosity mechanism assumed, but depends
on the inner torque boundary condition (\ref{eqn3}) (see also \citet{gammie00} for the 
effects of varying the torque at $r_{\rm {s}}$). Thus, the system of equations~(\ref{eqn1}), 
(\ref{eqn2}), (\ref{eqn4}) and~(\ref{eqn5}) decouples from the energy balance equation. 

Equation~(\ref{eqn4}) can be solved to give $H$ as 
\begin{equation}
H=\frac{B^2}{2\pi\Sigma\Omega_{\rm {K}}^2}\label{eqn6}\mbox{.}
\end{equation}
With the prescription for the viscous stress, equation~(\ref{eqn5}), the 
angular momentum conservation equation (\ref{eqn2}) becomes
\begin{equation}
\alpha\frac{B^4}{2\pi\Sigma}={\dot M}\left(1-\zeta\sqrt{\frac{r_{\rm {s}}}{r}}\right)
\Omega_{\rm {K}}^3\label{eqn7}\mbox{.}
\end{equation}

It is remarkable that the value of $\alpha$ can be constrained in our model.
In the framework of the local approach all the work done by non-gravitational 
forces on a patch of the disc is reduced to the work done by viscous stress $f_{\phi}$.
Physically, this work results from the action of turbulent and Maxwell stresses. 
Heating occurs from dissipation of supersonic 
turbulence. The rate of such heating 
can be expressed through 
the kinematic viscosity coefficient $\nu$ in the usual way (heating per unit volume)
\begin{equation}
\rho T \frac{ds}{dt}=\frac{f_{\phi}^2}{\nu \rho}\mbox{,}\label{eqn8}
\end{equation}
where $T$ is the average temperature of gas inside the disc, $s$ is the entropy per unit
mass (e.g., see in \citet{shapiro83}). Viscous stress, $f_{\phi}$, for Keplerian
shear is 
\begin{equation}
f_{\phi}=\frac{3}{2}\nu \rho \Omega_{\rm {K}}\mbox{.}\label{eqn9}
\end{equation}
Comparing this to the alternative expression~(\ref{eqn5}) for $f_{\phi}$ we 
see that $\alpha$ and $\nu$ are related by 
\begin{equation}
\nu=\alpha \frac{2}{3\Omega_{\rm {K}}} v_{\rm {A}}^2\mbox{.}\label{eqn10}
\end{equation}
Using expressions~(\ref{eqn9}) and~(\ref{eqn10}) to substitute in equation~(\ref{eqn8}) 
one can obtain the rate of heating per unit area of the disc $Q$ as follows
\begin{equation}
Q=2H \rho T \frac{ds}{dt}= 3\alpha \rho v_{\rm {A}}^3\mbox{.}\label{eqn11}
\end{equation}
One can also rewrite $Q$ as 
\begin{equation}
Q=\frac{3}{4\pi} \frac{GM{\dot M}}{r^3}\left(1-\zeta\sqrt{\frac{r_{\rm {s}}}{r}}\right)
\label{eqn11a}
\end{equation}
the same expression familiar in the standard disc model (e.g., see \citet{shapiro83}).
The total energy density of the turbulent pulsations and magnetic fields is 
$\displaystyle \frac{B^2}{8\pi}+\frac{1}{2}\rho v_{\rm {t}}^2 = \rho v_{\rm {A}}^2$ 
under the assumption of equipartition. In the stationary state, this turbulent
energy dissipates with the rate $Q$. Therefore, the characteristic time of 
the dissipation of the turbulence is 
\begin{equation}
\tau_{\rm {A}} = \frac{2H \rho v_{\rm {A}}^2}{Q} = 
\frac{2}{3\Omega_{\rm {K}} \alpha}\mbox{.}\label{eqn12}
\end{equation}

On the other hand, dissipation of the supersonic turbulence 
occurs on the time-scale $\tau_{\rm {t}}$ of the flow crossing the  
largest flow coherence size $l_{\rm {t}}$ of the 
turbulence.
Direct dissipation at the shock fronts dominates the turbulent 
cascade of energy down to the microscopic resistive scale and leads to 
the enhanced rate of the dissipation. The question of the dissipation 
of supersonic MHD turbulence has been studied in connection with the 
interstellar turbulence, which is observed to be highly supersonic. 
Direct numerical simulations of both steady-state 
driven and freely decaying MHD turbulence (\citealt{stone98}; \citealt{stone99}; 
\citealt*{ostriker01})
all confirm this picture. Even if initially the motion is set up 
to be incompressible in the numerical experiments, shocks develop 
rapidly due to the non-linear conversion of the waves and the dissipation 
becomes dominated by the dissipation on shocks.
This dissipation time is $\tau_{\rm {t}} = l_{\rm {t}}/ v_{\rm {t}}$. 
The coefficient of turbulent viscosity $\nu$
is also related to the largest flow coherence size and turbulent velocity as
\[
\nu=\frac{1}{3} l_{\rm {t}} v_{\rm {t}}\mbox{.}
\]
We already know that $v_{\rm {t}} = v_{\rm {A}}$ from 
equipartition of magnetic and kinetic energies
in turbulence. Also, $\nu$ is expressed through $\alpha$ by relation~(\ref{eqn10}).
This allows to estimate the largest flow coherence size of the turbulence
\begin{equation}
l_{\rm {t}}=2\alpha H\mbox{.}\label{eqn13}
\end{equation}
Consequently, the flow crossing time of the largest coherence size becomes
$\displaystyle \tau_{\rm {t}} = \frac{2\alpha}{\Omega_{\rm {K}}}$. 
It should be that $\tau_{\rm {t}}=\tau_{\rm {A}}$
with $\tau_{\rm {A}}$ given by expression~(\ref{eqn12}). This is possible only when 
$\alpha=1/\sqrt{3}$. Due to approximate nature of all calculations which 
lead us to this value for $\alpha$, it is only meaningful to state that $\alpha$ 
should be of order of $1$. We assume $\alpha=1$ for all further estimates. 
For $\alpha\approx 1$ the largest flow coherence size of the turbulence 
becomes of order of the 
thickness of the disc $l_{\rm {t}} \approx 2H$, and the turbulent viscosity coefficient 
takes its largest possible value $\nu\approx H v_{\rm {A}}$, still compatible with the 
local viscous description of the disc. The dissipation time-scale for the 
magnetic turbulence is $\tau_{\rm {A}} \approx 1/\Omega_{\rm {K}}$.
It is very probable that such large scale 
turbulence will lead to the buoyant rising of the magnetic field loops into the 
corona, subsequent shearing by the differential rotation and reconnection of 
the loops. This can result in the formation of the hot corona or acceleration
of particles to relativistic energies \citep{field93a, field93b}. 
The formation of a magnetized 
corona and the emission spectrum from the corona are important, 
however, here we focus on the disc.

The free parameters are ${\dot M}$ and $\zeta$. 
Also, we need to specify one more physical condition, 
since the dependence of $B$ on radius 
in equation~(\ref{eqn7}) is undetermined. 
Such a condition should come from the modelling of supersonic turbulence. 
Lacking a detailed model, we assume that the radial dependence of
the vertically averaged magnetic field in the disc is the power law 
\begin{equation}
B=B_{10} \left(\frac{r}{10 r_{\rm {g}}}\right)^{-\delta}\mbox{,}\label{eqn14}
\end{equation} 
where $B_{10}$ is the strength of the magnetic field at $10 r_{\rm {g}}$, $\delta > 0$ is
some constant. Accretion rate ${\dot M}$ can be related to the total luminosity 
of the disc $L$ and the radiated fraction $\epsilon$ of the rest mass accretion flux 
${\dot M} c^2$.
At the luminosity of an AGN 
\begin{equation}
L= 1.3\times 10^{46} l_{\rm {E}} M_8\,
\mbox{erg s}^{-1}\label{eqn15}\mbox{,}
\end{equation}
the  mass flux is
\begin{eqnarray}
&& {\dot M}=0.23\,{\rm M}_{\sun}\mbox{yr}^{-1}\left(\frac{l_{\rm {E}}}{\epsilon}\right)
M_8 = \nonumber\\
&& 1.4\times 10^{25}\,\mbox{g s}^{-1}
\left(\frac{l_{\rm {E}}}{\epsilon}\right) M_8
\label{eqn16}\mbox{.}
\end{eqnarray}
Here we denote the ratio of disc luminosity to the Eddington luminosity 
by $l_{\rm {E}}=L/L_{\rm {edd}}$. 
The value of $\epsilon$ is determined by the inner boundary condition at $r=r_{\rm {s}}$. 
Typically, $\epsilon \approx 0.1$.
Using such parametrization and the expression~(\ref{eqn14}) for the magnetic field,
one can obtain the following radial profiles of $H$, $\Sigma$ and $\rho$ from 
equations~(\ref{eqn6}) and~(\ref{eqn7})
\begin{eqnarray}
&& \frac{H}{r_{\rm {g}}}= \frac{\Omega_{\rm {K}} {\dot M} 
\mathcal{G}}{\alpha B^2 r_{\rm {g}}} =  
2.1\times 10^{-1}\, \frac{l_{\rm {E}}}{2\epsilon}
\left(\frac{B_{10}}{10^4\,\mbox{G}}
\right)^{-2} \times \nonumber\\
&& M_8^{-1} {\mathcal{G}} \left(\frac{r}{10 r_{\rm {g}}}\right)^{2\delta-3/2}
\label{eqn17}\mbox{,} \\
&& \Sigma = \frac{\alpha B^4}{2\pi \Omega_{\rm {K}}^3 {\dot M} {\mathcal{G}}} =   
5.1\times 10^{3}\,\mbox{g cm}^{-2} 
\left(\frac{l_{\rm {E}}}{2\epsilon}\right)^{-1} \times \nonumber\\
&& \left(\frac{B_{10}}{10^4\,\mbox{G}}
\right)^4 M_8^2 {\mathcal{G}}^{-1} \left(\frac{r}{10 r_{\rm {g}}}\right)^{9/2-4\delta}
\label{eqn18}\mbox{,} \\
&& \rho= \frac{\alpha^2 B^6}{4\pi \Omega_{\rm {K}}^4 {\dot M}^2 {\mathcal{G}}^2} = 
4\times 10^{-10}\,\mbox{g cm}^{-3}\,
\left(\frac{l_{\rm {E}}}{2\epsilon}\right)^{-2} \times \nonumber\\
&& \left(\frac{B_{10}}{10^4\,\mbox{G}}\right)^6 {\mathcal{G}}^{-2} M_8^2 
\left(\frac{r}{10 r_{\rm {g}}}\right)^{6-6\delta}         
\label{eqn19}\mbox{,}   
\end{eqnarray}
where $\displaystyle \mathcal{G}=1-\zeta\sqrt{\frac{r_{\rm {s}}}{r}}$. 
We see that the disc becomes geometrically thicker and less dense when magnetic 
field decreases: weaker magnetic field leads to
weaker tangential stress (equation~(\ref{eqn5})); $H$ increases in order to 
accomodate constant angular momentum flux for the same ${\dot M}$ such that
$H\propto B_{10}^{-2}$ (equation~(\ref{eqn2})); $\Sigma$ and $\rho$ should 
decrease strongly, $\Sigma \propto B_{10}^4$ and $\rho\propto B_{10}^6$, in order
to ensure vertical equilibrium with larger $H$ and less pressure support
from $B^2$ (equation~(\ref{eqn6})); radial inflow velocity increases as 
$v_r\propto B_{10}^{-4}$ to ensure constant mass flux.

Let us now summarize the similarities and differences between our 
model and the standard Shakura-Sunyaev model. If we replace the thermal
pressure in the standard model by the sum of the magnetic and turbulent 
pressures, the equations for mass conservation~(\ref{eqn1}), 
angular momentum conservation~(\ref{eqn2}), the viscosity 
prescription~(\ref{eqn5}) and vertical pressure 
support~(\ref{eqn4}) are the same as in the standard model. The pressure in 
the standard model is determined by the rate of the cooling of the 
disc, while the $\alpha$ coefficient can be an arbitrary function of 
$r$, $\alpha(r)<1$. In our model we have the magnetic pressure
unspecified in its radial dependence as 
soon as it exceeds the thermal pressure, but 
$\alpha\approx 1$ for all $r$. The  latter results from  
the much faster dissipation of supersonic turbulence than  
subsonic turbulence assumed in the standard model. Both our model
and the standard model have only one undetermined function of radius, 
($\alpha(r)$ in the standard model and $B(r)$ 
in our model). The determination of this free function would eventually 
come from detailed modelling of the MHD turbulence.

\section{Estimates of the Disc Parameters}\label{sec3}

Now let us use the solution for the disc structure provided by equations~(\ref{eqn6}),
(\ref{eqn7}) and~(\ref{eqn14}) and obtain constraints on free parameters of
the model, such that our model of thin magnetized accretion disc is self-consistent. 
Using equations~(\ref{eqn7}) and~(\ref{eqn6}) to substitute for ${\dot M}$ in 
equation~(\ref{eqn1}) we can express the radial inflow velocity
as 
\begin{equation}
\frac{v_r}{v_{\rm {K}}}=\frac{\alpha}{\mathcal{G}}\frac{H^2}{r^2}\mbox{.}\label{eqn3.1}
\end{equation}
The factor ${\mathcal{G}}$ 
vanishes at $r=r_{\rm {s}}$ for $\zeta=1$ and so the surface density $\Sigma$ of the 
disc diverges near $r=r_{\rm {s}}$. The same divergence also occurs in standard disc 
model \citep{shakura73}. 
In reality, of course, viscous torque does not vanish
at $r=r_{\rm {s}}$, $\zeta \neq 1$ but close to $1$ and there is no divergence of 
$\Sigma$ at $r=r_{\rm {s}}$. 
Determining $\zeta$ would require a full general relativistic 
treatment of the accretion flow close to the black hole and is beyond the 
immediate scope of this work. Only parts of the disc close to $r_{\rm {s}}$  
are sensitive to the exact value of $\zeta$. For $r>r_{\rm {s}}$,
the disc structure is approximated well with the formulae in section~\ref{sec2} 
for $\zeta=1$. Provided that $r> r_{\rm {s}}$ and $\alpha\leq 1$ one can see that the 
radial inflow velocity for a thin disc ($H\ll r$) is always a small fraction 
of the Alfv\'en velocity which, in turn, is a small fraction of the Keplerian 
velocity. Therefore, the dissipation time-scale $\tau_{\rm {A}}$ 
is always much shorter than the radial inflow time-scale. In the 
stationary case this means that the advective terms in the energy balance equation
can be always neglected. Energy balance becomes local: the rate of gas heating 
$Q$ should be balanced by the cooling due to radiative processes. 
Now we consider the physics of radiative cooling which determines the disc
temperature.

A necessary condition for the existence of a magnetically dominated disc is that
the vertical escape time for radiation must be shorter than $\tau_{\rm {A}}$, 
so that the energy density of
radiation, $\varepsilon_{\rm {r}}=a T^4$, remains smaller than the energy density of the 
magnetic field $\displaystyle \varepsilon_{\rm {B}}=\frac{B^2}{8\pi}$. For optically thin, 
geometrically thin discs this condition is always satisfied since the inverse of 
escape time $c/H \gg \Omega_{\rm {K}}$. As we will see, Thomson scattering
is the dominant source of opacity in most cases of optically thick discs. 
The average time it takes for a photon to escape out of optically thick disc with 
optical depth $\tau\gg 1$ is $\tau H/c$. For Thomson scattering $\tau= \tau_{\rm {c}} =
H n_{\rm {e}} \sigma_{\rm {T}} = \kappa_{\rm {T}} \Sigma/2$, where 
$n_{\rm {e}}$ is the number density of free electrons, 
$\sigma_{\rm {T}}$ is the Thomson cross section 
and $\kappa_{\rm {T}}=0.4\,\mbox{cm}^2\mbox{g}^{-1}$ 
is the Thomson scattering opacity. For simplicity
we assume the composition of the disc to be completely ionized hydrogen. 
Then, $n_{\rm {e}}$ is equal to the number density of protons in the disc, $n$. The necessary
condition now becomes 
\begin{equation}
\frac{H^2}{c} n \sigma_{\rm {T}} < \frac{1}{\Omega_{\rm {K}}}\mbox{.}\label{eqn3.2} 
\end{equation}
Using equations~(\ref{eqn6}) and~(\ref{eqn7}) one can rewrite the condition~(\ref{eqn3.2})
as 
\begin{equation}
\frac{l_{\rm {E}}}{2\epsilon}\frac{r_{\rm {g}}}{H} {\mathcal{G}}^2 < 1\mbox{.}\label{eqn3.3}
\end{equation}
We express this and all the subsequent conditions in terms of free parameters of 
the model: $B_{10}$,  $\delta$, $l_{\rm {E}}/\epsilon$, and $M_8$. Using the 
expression~(\ref{eqn17}) for $H$, the necessary condition~(\ref{eqn3.3}) becomes
\begin{equation}
4.7\times \left(\frac{B_{10}}{10^4\,\mbox{G}}
\right)^{2} M_8 {\mathcal{G}} \left(\frac{r}{10 r_{\rm {g}}}\right)^{3/2-2\delta} \ll 1
\mbox{.}\label{eqn3.4}
\end{equation}

The condition for the disc to be optically thick for Thomson scattering 
$\tau_{\rm {c}} \gg 1$ becomes
\begin{eqnarray}
&& 2.0\times 10^{3}\, 
\left(\frac{l_{\rm {E}}}{2\epsilon}\right)^{-1} \left(\frac{B_{10}}{10^4\,\mbox{G}}
\right)^4 \times \nonumber\\
&& M_8^2 {\mathcal{G}}^{-1} \left(\frac{r}{10 r_{\rm {g}}}\right)^{9/2-4\delta}
\gg 1\mbox{.}\label{eqn3.5}
\end{eqnarray}
In an optically thick disc, radiation is transported by turbulent motions and 
radiative diffusion. The characteristic time-scale for the turbulent transport of 
radiation to the surface of the disc is $\sim 1/\Omega_{\rm {K}}$.  
When condition~(\ref{eqn3.2}) is satisfied, the diffusion of radiation dominates the 
advection due to turbulent motions. Thus, we can neglect turbulent convective 
transport of radiation for any values of parameters, whenever steady state 
magnetically dominated model of the accretion disc is considered. 
In the steady state, the radiation flux from 
each surface of the disc must be equal to $Q/2$ with the dissipation rate $Q$ given 
by expressions~(\ref{eqn11}) or~(\ref{eqn11a}). The effective temperature of the 
escaping radiation flux is determined by this dissipation rate $Q$ as 
\begin{eqnarray}
&& T_{\rm {eff}}=\left(\frac{2Q}{ac}\right)^{1/4}=7.5\times 10^4\,\mbox{K}
\left(\frac{l_{\rm {E}}}{2\epsilon}\right)^{1/4}\times \nonumber\\
&& \left(\frac{r}{10 r_{\rm {g}}}\right)^{-3/4}{\mathcal{G}}^{1/4} M_8^{-1/4}
\mbox{,}\label{eqn3.6}
\end{eqnarray}  
and is the same in our model as in the standard disc model. 
True absorption 
of photons in free-free transitions also occur in the disc. For an approximate 
estimate of the radiative conditions in the disc we consider Rosseland mean of the
free-free absorption opacity
\begin{equation}
{\bar \kappa}_{\rm {ff}}= 6.4\times 10^{22}\,\mbox{cm}^2 \mbox{g}^{-1} \rho T^{-7/2}\mbox{,}
\label{eqn3.7}
\end{equation}
where $\rho$ is expressed in $\mbox{g cm}^{-3}$ and $T$ is expressed in 
$^{\degr }\mbox{K}$. When the effective optical thickness of the disc 
$2{\bar \tau}_{\ast}=
\Sigma\sqrt{(\kappa_{\rm {T}}+{\bar \kappa}_{\rm {ff}}){\bar \kappa}_{\rm {ff}}}$ 
is large, then 
local thermal equilibrium is established in the disc and the radiative flux is 
described by diffusion approximation. Only in the thin surface layer at a distance from the 
surface less then the thermalization optical depth ${\bar \tau_{\ast}}\approx 1$, the 
spectrum of radiation deviates substantially from that of the black body 
(see section~\ref{sec4} below). The solution of the diffusive radiation transport gives
the usual result relating the temperature at the midplane of the disc, $T_{\rm {mpd}}$, 
with the effective temperature at the surface
\citep{shakura73, krolik99}
\begin{eqnarray}
&& T_{\rm {mpd}} \approx T_{\rm {eff}}\left(\frac{3 \kappa_{\rm {T}} 
\Sigma}{16}\right)^{1/4} =
3.3\times 10^5\,\mbox{K} \times \nonumber\\
&& \left(\frac{B_{10}}{10^4\,\mbox{G}}
\right) M_8^{1/4}\left(\frac{r}{10 r_{\rm {g}}}\right)^{3/8-\delta}
\mbox{.}\label{eqn3.8}
\end{eqnarray}
The averaged temperature in the disc is close to $T_{\rm {mpd}}$ with the actual profile 
being determined by the details of the vertical dependence of the dissipation 
rate in shocks. As in the case  of the inner radiation dominated part of the 
standard disc, $T_{\rm {mpd}}$ does not depend on the accretion rate. However, it is 
directly proportional to the value of the magnetic field and has radial 
dependence governed by the $\delta$.

The dominance of the magnetic and turbulent energy compared to the energy density 
of radiation is expressed as $aT_{\rm {mpd}}^4 \ll \rho v_{\rm {A}}^2$. 
One can substitute here for $T_{\rm {mpd}}$
from equations~(\ref{eqn3.8}) and~(\ref{eqn3.6}). Heating rate $Q$ is given by 
equation~(\ref{eqn11}).  After using 
equations~(\ref{eqn17}--\ref{eqn19}) to manipulate with $\Sigma$, $B$ and $H$, 
one can reduce the condition of magnetic pressure dominance over the 
radiation pressure to be  
\begin{equation}
\frac{9\alpha}{4c}\sigma_{\rm {T}} n H^2 \ll \frac{1}{\Omega_{\rm {K}}}\mbox{.}\label{eqn3.9}
\end{equation}
This differs only by a factor of order unity from the necessary 
condition for the escape of radiation, equation~(\ref{eqn3.2}). 
The assertion that the condition~(\ref{eqn3.4})
implies smallness of radiation pressure is thus confirmed by direct calculation.
The next condition we need to impose is that the gas pressure is
small compared to the magnetic pressure and turbulent stresses, which is equivalent 
to the statement that the turbulence is highly supersonic. For a thermalized 
radiation field, this condition is $nkT_{\rm {mpd}} \ll \rho v_{\rm {A}}^2$. 
Using expressions~(\ref{eqn3.8}) and~(\ref{eqn3.6}) for the temperature and 
substituting for the density and $H$ from expressions~(\ref{eqn17}--\ref{eqn19}) 
one can express the condition $nkT_{\rm {mpd}} \ll \rho v_{\rm {A}}^2$ as 
\begin{equation}
\left(\frac{9 k^4 \sigma_{\rm {T}}}{32\pi^2 ac m_{\rm {p}}^5}\right)^{1/2} \alpha^{9/2}
\Omega_{\rm {K}}^{-17/2} B^{10} {\dot M}^{-4} {\mathcal{G}}^{-4} \ll 1 
\mbox{.}\label{eqn3.10}
\end{equation}
Substituting for $\Omega_{\rm {K}}$, $B$, and ${\dot M}$ in~(\ref{eqn3.10}) 
and taking the $-1/5$ power we obtain
\begin{eqnarray}
&& 13\times M_8^{-9/10}\left(\frac{r}{10 r_{\rm {g}}}\right)^{-51/20+2\delta}
{\mathcal{G}}^{4/5} \times \nonumber\\
&& \left(\frac{l_{\rm {E}}}{2\epsilon}\right)^{4/5}
\left(\frac{B_{10}}{10^4\,\mbox{G}}\right)^{-2} \gg 1
\mbox{.}\label{eqn3.11}
\end{eqnarray}

Because the free-free opacity ${\bar \kappa}_{\rm {ff}}$ strongly depends on temperature 
and density (equation~(\ref{eqn3.7})), the vertical density and temperature 
distributions are needed to
evaluate ${\bar\tau}_{\ast}$. Simulations in \citet{miller00} show a sharp 
density drop off by two orders of magnitude at the surfaces of the disc (see figure~11
in \citet{miller00}). Inside the slab bounded by this density drop off the 
magnetic field and kinetic energy are in approximate equipartition. 
For the purposes of calculating effective optical depth for absorption we 
approximate the density profile between $z=-H$ and $z=H$ inside the disc as constant
and assume zero density at the disc surfaces at $z=\pm H$. Modest 
variations of magnetic field and density across the disc height observed in numerical
simulations support this and also suggest the assumption of a uniform 
turbulence dissipation rate across the thickness.
With these approximations, the solution
for diffusive radiation transport in the vertical direction $z$ can be written
\citep{shakura73,krolik99}
\begin{equation}
T^4=T_{\rm {mpd}}^4 \left[\frac{8}{3\tau_{\rm {c}}}+\left(1-\frac{z}{H}\right)\right]
\mbox{.}\label{eqn3.12}
\end{equation} 
The Eddington approximation near the surface $z=H$ of the disc was used to obtain this 
temperature distribution, and $T_{\rm {eff}}=\sqrt[4]{2} \left. T\right|_{z=H}$ according
to the solution~(\ref{eqn3.12}). This solution is not valid near the surface 
of the disc in the region dominated by Compton scattering but gives $T$ in the bulk 
of the disc if ${\bar\tau}_{\ast} \gg 1$. Using the expression~(\ref{eqn3.12}) for $T$ one 
has for the optical depth of the disc for free-free absorption
\begin{eqnarray}
&& {\bar\tau}_{\rm {ff}}=\int_0^H \rho \kappa_{\rm {ff}}\,dz = 6.4 \times 10^{22}
\,\rho^2 T_{\rm {mpd}}^{-7/2} H \times \nonumber\\
&& \int_0^1 \left(1-\xi+\frac{8}{3\tau_{\rm {c}}}\right)^{-7/8}
\,d\xi  \mbox{,}\label{eqn3.13}
\end{eqnarray}
where $\xi=z/H$, $\rho$ and $T$ are expressed in $\mbox{g cm}^{-3}$ and $^{\degr }\mbox{K}$.
The value of the integral over $\xi$ determines how much the actual 
value ${\bar\tau}_{\rm {ff}}$ is larger than the value of ${\bar\tau}_{\rm {ff}}$ 
obtained if one assumes 
$T=T_{\rm {mpd}}$ everywhere in the disc. The $\xi$ integral in equation~(\ref{eqn3.13})
is calculated to be 
\begin{eqnarray}
&& \int_0^1 \left(1-\xi+\frac{8}{3\tau_{\rm {c}}}\right)^{-7/8}\,d\xi = 
\nonumber\\
&& 8\left[\left(1+\frac{8}{3\tau_{\rm {c}}}\right)^{1/8}-
\left(\frac{8}{3\tau_{\rm {c}}}\right)^{1/8}\right] \mbox{.}
\nonumber
\end{eqnarray}
Because $\tau_{\rm {c}} \gg 1$ the number in square parenthesis is close to $1$, so we can omit
it and obtain the final result for ${\bar\tau}_{\rm {ff}}$ 
\begin{equation}
{\bar\tau}_{\rm {ff}}= 5 \times 10^{23}\,\rho^2 T_{\rm {mpd}}^{-7/2} H \mbox{.}\label{eqn3.14}
\end{equation}
When one substitutes for $T_{\rm {mpd}}$, $H$, and $\rho$ their expressions through the 
magnetic field and the accretion rate, the expression~(\ref{eqn3.14}) becomes
\begin{eqnarray}
&& {\bar\tau}_{\rm {ff}}=8.6\times 10^{-2}\,M_8^{25/8}
\left(\frac{r}{10 r_{\rm {g}}}\right)^{147/16-13\delta/2} 
{\mathcal{G}}^{-3} \times \nonumber\\
&& \left(\frac{l_{\rm {E}}}{2\epsilon}\right)^{-3} 
\left(\frac{B_{10}}{10^4\,\mbox{G}}\right)^{13/2}\mbox{.}\label{eqn3.15}
\end{eqnarray} 
Generally, for $\delta \approx 1$, $M_8\la 1$, $l_{\rm {E}}/\epsilon \sim 0.1$, $B_{10} 
\la 10^4\,\mbox{G}$, and $r \sim 10 r_{\rm {g}}$, ${\bar\tau}_{\rm {ff}}\sim 1$. 
However, because of steep dependence of ${\bar\tau}_{\rm {ff}}$ on $l_{\rm {E}}$, $M_8$,
and $B_{10}$, the value of ${\bar\tau}_{\rm {ff}}$ can become large for lower accretion rates,
more massive black holes, and stronger magnetic fields. The ratio of ${\bar\tau}_{\rm {ff}}$ 
to $\tau_{\rm {c}}$ is
\begin{eqnarray}
&& \frac{{\bar\tau}_{\rm {ff}}}{\tau_{\rm {c}}}=4.2\times 10^{-5}\,M_8^{9/8}
\left(\frac{r}{10 r_{\rm {g}}}\right)^{75/16-5\delta/2} 
{\mathcal{G}}^{-2} \times \nonumber\\
&& \left(\frac{l_{\rm {E}}}{2\epsilon}\right)^{-2} 
\left(\frac{B_{10}}{10^4\,\mbox{G}}\right)^{5/2}
\mbox{.}\label{eqn3.16}
\end{eqnarray}
We see that for the typical values of the parameters above the ratio 
${\bar\tau}_{\rm {ff}}/\tau_{\rm {c}} \sim 10^{-2}$ 
but free-free optical depth can exceed Compton
scattering optical depth for smaller values of $l_{\rm {E}}$ and at larger radii. When 
${\bar\tau}_{\rm {ff}} \ll \tau_{\rm {c}}$ the effective optical thickness of the disc becomes
${\bar\tau}_{\ast}=\sqrt{{\bar\tau}_{\rm {ff}}\tau_{\rm {c}}}$ or
\begin{eqnarray}
&& {\bar\tau}_{\ast}=13\times M_8^{41/16} 
\left(\frac{r}{10 r_{\rm {g}}}\right)^{219/32-21\delta/4} 
{\mathcal{G}}^{-2} \times\nonumber\\
&& \left(\frac{l_{\rm {E}}}{2\epsilon}\right)^{-2} \left(\frac{B_{10}}{10^4\,\mbox{G}}
\right)^{21/4}
\mbox{.}\label{eqn3.17}
\end{eqnarray}
When 
${\bar\tau}_{\rm {ff}} \gg \tau_{\rm {c}}$ the effective optical thickness is equal to 
${\bar\tau}_{\rm {ff}}$.  

\begin{figure}
\includegraphics[width=\textwidth]{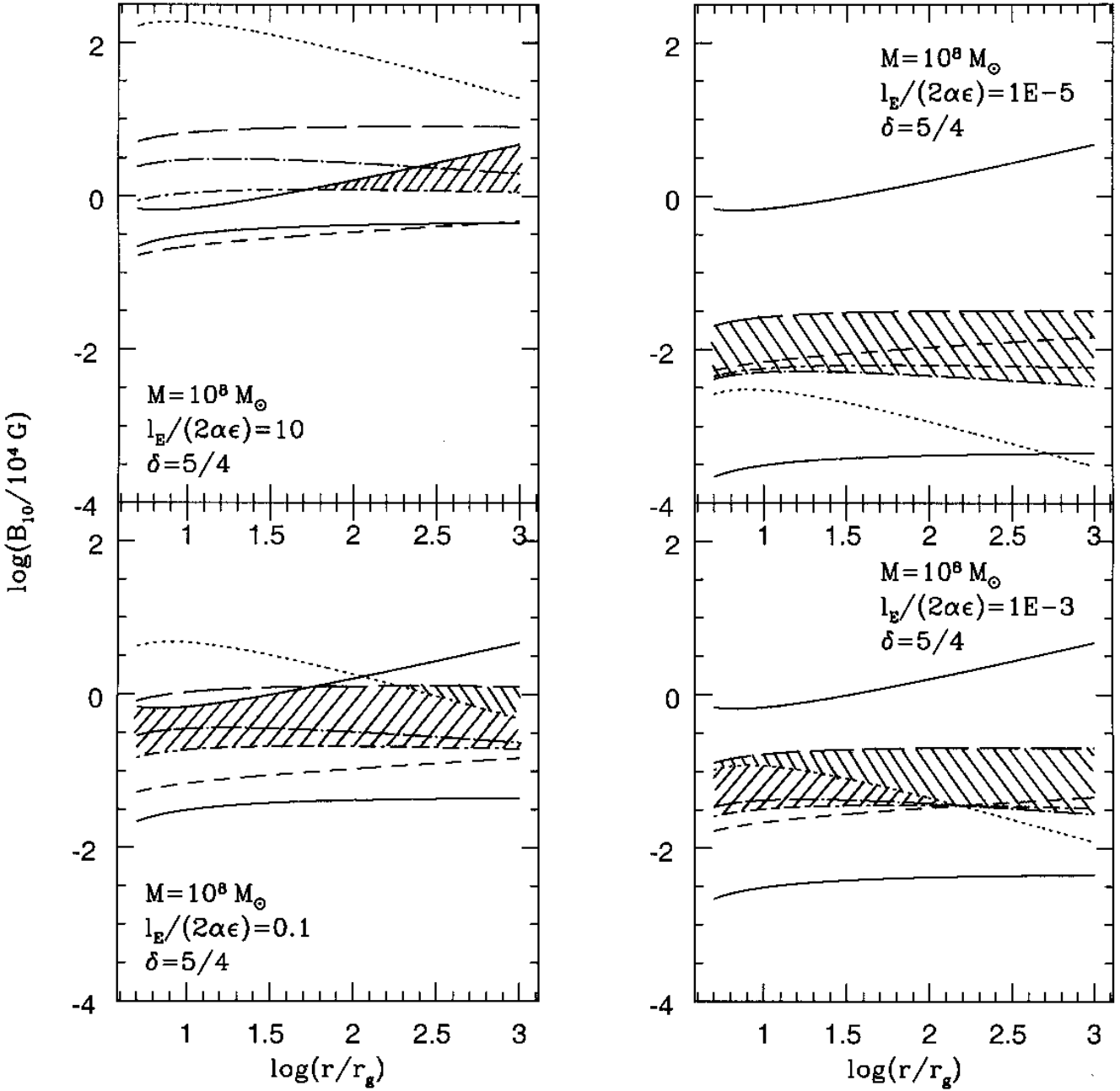}
\caption{
Plots of conditions when our model is valid. 
Plots are for $M=10^8\,{\rm M}_{\sun}$ and $\delta=5/4$. 
On each plot values of radius extend from $5 r_{\rm {g}}$ to 
$1000 r_{\rm {g}}$. The panels differ by values of $l_{\rm {E}}/(2\alpha\epsilon)$. 
On each panel, any given 
model of the disc is represented with a horizontal line $B_{10}=\mbox{constant}$. 
Filled areas indicate regions where a magnetically dominated geometrically thin
and optically thick
disc can exist. The difference in filling represents different types of spectra
emitted locally from the surface of the disc: the regions with black body spectra 
are filled with lines in top-left to bottom-right direction and 
the regions with 
modified black body spectra are filled with lines in bottom-left to top-right 
direction. 
There are seven types of lines on each panel: (1) upper solid 
line curved upward on each plot bounds the region where radiation pressure 
is small compared to magnetic pressure and turbulent stress (below the curve);
(2) lower solid line curved downward on each plot bounds the region where the 
disc is thin, i.e. $H< r$ (above the curve); (3) long-dashed line bounds the 
region where where thermal gas pressure is small compared to magnetic pressure 
and turbulent stress (below the curve); (4) short-dashed line separates the 
regions where the disc Thomson optical depth $\tau_{\rm {c}} < 1$ (above the line) and
$\tau_{\rm {c}} > 1$ (below the line); (5) long-dashed and dotted line separates the 
region where the disc free-free optical depth ${\bar\tau}_{\rm {ff}} > 1$ 
(above the line) and ${\bar\tau}_{\rm {ff}} < 1$ (below the line);
(6) short-dashed and dotted line separates the 
regions where the disc effective optical depth $\sqrt{{\bar\tau}_{\rm {ff}}\tau_{\rm {c}}} > 1$ 
(above the line) and $\sqrt{{\bar\tau}_{\rm {ff}}\tau_{\rm {c}}} < 1$ (below the line);
(7) dotted line separates the regions where the disc free-free optical depth
exceeds Thomson optical depth, ${\bar\tau}_{\rm {ff}} > \tau_{\rm {c}}$, (above the line)
from where ${\bar\tau}_{\rm {ff}} < \tau_{\rm {c}}$, (below the line). 
}
\label{fig_m8d54}
\end{figure}

\begin{figure}
\includegraphics[width=\textwidth]{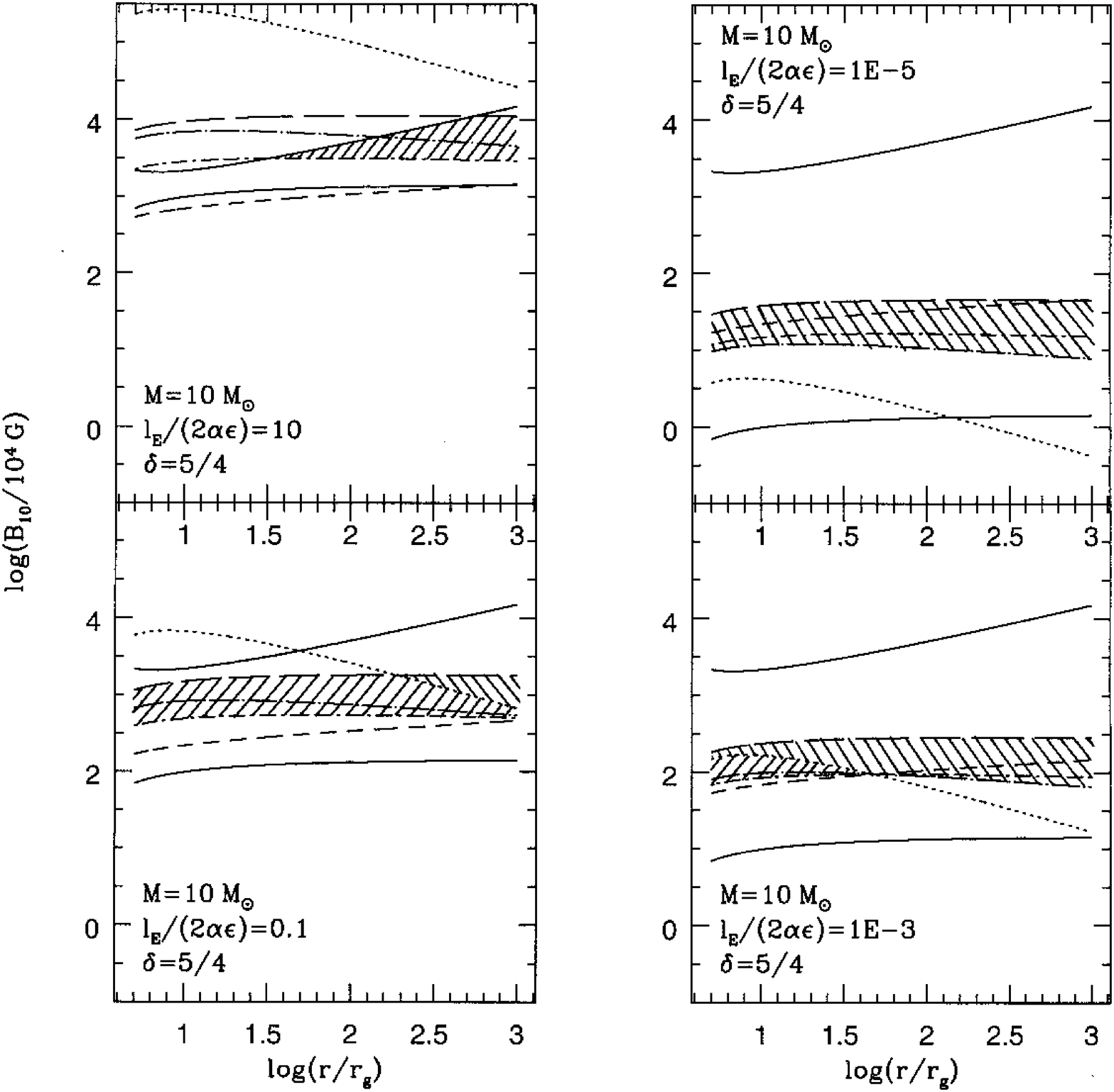}
\caption{
Plots of conditions when our model is valid. Plots are done in the plane of 
$B_{10}$ and $r/r_{\rm {g}}$ for $M=10\,{\rm M}_{\sun}$ and $\delta=5/4$. All notations
are the same as in Fig.~\protect\ref{fig_m8d54}.
}
\label{fig_m1d54}
\end{figure}

Finally, the ratio of the disc semi-thickness $H$ to the 
radius $r$ using (\ref{eqn17}) is
\begin{equation}
\frac{H}{r}= 2.1\times 10^{-2}\, \frac{l_{\rm {E}}}{2\epsilon}
\left(\frac{B_{10}}{10^4\,\mbox{G}}
\right)^{-2} M_8^{-1} {\mathcal{G}} \left(\frac{r}{10 r_{\rm {g}}}\right)^{2\delta-5/2}
\label{eqn3.18}\mbox{.}
\end{equation}
Now we summarize 
conditions when our model is valid: 
\begin{enumerate}
\item The necessary condition~(\ref{eqn3.4}), which is also the condition for
the dominance of the magnetic pressure over the radiation pressure.
\item The condition~(\ref{eqn3.11}) for the dominance of the magnetic 
pressure over the thermal pressure of gas in thermalized optically thick disc. 
\item The condition for the disc to be optically thick. This is either 
condition~(\ref{eqn3.5}) $\tau_{\rm {c}} \gg 1$ or the condition that ${\bar\tau}_{\rm {ff}}$ 
given by expression~(\ref{eqn3.15}) is greater than $1$. 
\item The condition that the gas and radiation inside the disc are in 
local thermal equilibrium, ${\bar\tau}_{\ast} \gg 1$, where  
${\bar\tau}_{\ast}={\bar\tau}_{\rm {ff}}$ if ${\bar\tau}_{\rm {ff}}/\tau_{\rm {c}} > 1$, and  
${\bar\tau}_{\ast}$ is given by expression~(\ref{eqn3.17}) if 
${\bar\tau}_{\rm {ff}}/\tau_{\rm {c}} < 1$. 
\item $H/r \ll 1$. 
\end{enumerate}
We varied the parameters $l_{\rm {E}}/\epsilon$, $\delta$, $M_8$ 
to obtain the allowed region of our disc model
in the $B_{10}$, $r/r_{\rm {g}}$ plane, using the above five conditions. 
These plots are shown in 
Figs.~\ref{fig_m8d54} --~\ref{fig_m1d54}.

Depending on the power $\delta$ in the dependence of 
magnetic field $B\propto r^{-\delta}$, optically thick magnetically dominated
accretion discs can exist only at a limited interval of radii. 
For $\delta \approx 1$
(as on our plots for $\delta=5/4$) a thin magnetically dominated 
disc (shadowed regions on the plots) is possible for $5 r_{\rm {g}}<r<1000 r_{\rm {g}}$. 
The window for the strength of the magnetic field is not very wide:
about one order of magnitude or less. This window is narrower for low masses
of the central black hole and is wider for higher masses. 
The value $l_{\rm {E}}/(2\alpha\epsilon)=10$ corresponds to about the Eddington accretion 
rate for $\epsilon\approx 0.1$, and because of $\alpha \approx 1$ 
in our model (see section~2). Higher values of $l_{\rm {E}}/(2\alpha\epsilon)$ correspond
to higher accretion rates. Allowed values of the magnetic field are higher for
higher accretion rates.
The magnetic fields in the discs around higher mass black holes are smaller 
than in the discs around lower masses black holes as is temperature 
of the disc ($T_{\rm {mpd}}$) and the surface radiation flux.
For large luminosities ($l_{\rm {E}}/(2\alpha\epsilon)\ga 1$) the inner 
disc cannot be optically thick for true absorption but can be optically 
thick to free electron scattering. Comptonisation
becomes significant for in the inner regions at high 
luminosities (see section~\ref{sec4}). We leave consideration of 
Comptonised regimes for future work.

Physically, the limitations on the magnetic field strength can be understood as 
follows: suppose one decreases $B_{10}$ while keeping ${\dot M}$
constant. 
Then, $H \propto B_{10}^{-2}$ is increasing; 
$\Sigma \propto B_{10}^4$, and $\rho \propto
B_{10}^6$, both decreasing (equations (\ref{eqn17}--\ref{eqn19})).
Scattering opacity through the disc $\tau_{\rm {c}} \propto \Sigma$
strongly decreases, so the heat is transported to the surface faster and 
$T_{\rm {mpd}}\propto B_{10}$ decreases (equation~(\ref{eqn3.8})); 
thermal and radiation pressures decrease as $P_{\rm {th}}\propto B_{10}^7$
and $P_{\rm {rad}}\propto B_{10}^4$ respectively; 
plasma parameter $\beta$ decreases, so 
the disc becomes more magnetically dominated; ${\bar\tau}_{\rm {ff}}$ and 
${\bar\tau}_{\ast}$ both decrease since their decrease due to lower $\Sigma$  
overcomes the increase of the absorptive opacity from the drop of 
the temperature. Therefore, there exists only a limited interval of
$B_{10}$ such that $\beta<1$ still the disc is optically thick to true absorption.

\section{Radiation Spectra of Optically Thick Magnetically Dominated Disc}
\label{sec4}

\subsection{Calculation of spectra}
\label{subsec41}

Free-free, bound-free and cyclotron emission could contribute to 
the radiation spectrum. 
In Appendix~A we show that, because of the 
self-absorption in the dense disc, 
the total flux of cyclotron emission 
from the disc surface is negligibly small compared to the total radiated
power $Q$. 
This power is entirely due to free-free and bound-free 
radiative transitions.
Cyclotron and synchrotron emission can be important in the 
rarefied and strongly magnetized disc corona (\citealt{ikhsanov89};
\citealt{field93a}, \citealt{dimatteo97a}), but our focus here is on the disc.

We perform a simplified calculation of the emergent spectrum. We assume local
thermodynamic equilibrium and do not consider effects of the temperature change with
depth. This is justified when the spectrum is formed in thin layer near 
the disc surface. For simplicity we do not include the bound-free
contribution to the opacity. Free-free opacities 
are the dominant source of thermal absorption for $T \ga 10^5\,\mbox{K}$, so
our simplified spectrum is most relevant for smaller masses 
of the central black hole, for which the inner disc is hotter. Our goal here is to
capture the effect of the magnetic field on the shape of the spectrum. 
We consider only 
optically thick disc models with both $\tau_{\rm {c}}\gg 1$ and $\tau_\ast \gg 1$.
The electron scattering opacity $\kappa_{\rm {T}}$ does not depend on frequency in 
the non-relativistic limit, whereas free-free absorption opacity 
$\kappa_{\rm {ff}}(\nu)$ is a function of frequency:
\begin{equation}
\kappa_{\rm {ff}}(\nu)=1.5\times 10^{25}\,\mbox{cm}^2\mbox{g}^{-1}\rho T^{-7/2}
\frac{1-e^{-x}}{x^3}g(x,T)\label{eqn4.5}\mbox{,}
\end{equation} 
where we denote $\displaystyle x=\frac{h\nu}{kT}$. Thus, the parameters of
the disc model $B_{10}$ and $\delta$ will affect the spectrum by means 
of $\kappa_{\rm {ff}}$ dependence on $\rho$ and $T/T_{\rm {eff}}$. Gaunt factor $g(x,T)$ 
is slowly varying function of $x$ and $T$ between approximately $0.3$ and $5$
in wide range of frequencies and temperatures \citep{rybicki79}.    
Moreover, in the wide temperature interval $10^3\,\mbox{K}<T<10^8\,\mbox{K}$
Gaunt factor is between $0.5$ and $2$ for frequencies $0.1<x<10$ near the 
maximum of thermal emission. It is quite reasonable to set $g(x,T)=1$, 
which we do in all further calculations.
Note, that $\kappa_{\rm {ff}}$ behaves as $1/x^2$ for $x\ll 1$, so free-free absorption 
is always more important at lower frequencies than electron scattering. 

The energy transfer due to repeated scatterings (Comptonisation process) is 
characterized by Compton $y(\nu)$ parameter \citep{rybicki79}
\begin{equation}
y=\frac{4kT}{m_{\rm {e}} c^2}\,\mbox{Max}\left(\tau_{\rm {es}},
\tau_{\rm {es}}^2\right)\label{eqn4.6}
\mbox{,}
\end{equation}
where the optical depth for Thomson scattering $\tau_{\rm {es}}(\nu)$ 
must be measured from the effective 
absorption optical thickness $\tau_{\ast}(\nu)\sim 1$ and is given by (formula
[7.42] in \citet{rybicki79})
\begin{equation}
\tau_{\rm {es}}(\nu)\approx \left(\frac{\kappa_{\rm {T}}/
\kappa_{\rm {ff}}(\nu)}{1+\kappa_{\rm {ff}}(\nu)
/\kappa_{\rm {T}}}\right)^{1/2}\label{eqn4.7}\mbox{.}
\end{equation}
If $y(\nu)\ll 1$, photons do not change their initial frequency $\nu$ in the 
process of repeated scatterings before they escape the surface of the disc.
While if $y\ga 1$ the Comptonisation effects become significant. We calculated
values of $y$ using expressions~(\ref{eqn4.6}), (\ref{eqn4.7}) and~(\ref{eqn4.5}).
One can see that $y$ is always monotonically rising with frequency, therefore,
the Comptonisation effect at higher frequencies is always more significant 
than at lower frequencies. On the other hand, there is very little radiation
at $h\nu \geq 5 kT$ due to exponential cut off in thermal spectra.
It turns out that in most cases $y(\nu)\ll 1$ for optically thick disc models
and for $h\nu< 5kT$. 
Exceptions are the cases of high accretion rates $l_{\rm {E}}/\epsilon \ga 1$. 
In those cases, inner parts 
of the accretion disc ($r<30 r_{\rm {g}}$) can have $y\geq 1$ and Comptonisation 
will influence the highest frequencies of the disc spectrum. Ignoring these
exceptions, we did not take into account Comptonisation in the following
calculations and assume coherent scattering. This assumption has been checked 
a posteriori for self-consistency.

In the case of coherent scattering the approximate expression of the radiative
flux per unit surface of an optically thick medium is given by 
\citep{rybicki79}
\begin{equation}
F_{\nu}\approx\frac{4\pi h\nu^3/c^2}{\left(e^{h\nu/kT}-1\right)
\left(1+\sqrt{1+\kappa_{\rm {T}}/\kappa_{\rm {ff}}
(\nu,T)}\right)}\label{eqn4.8}\mbox{.}
\end{equation}
$F_{\nu}$ approaches black body spectrum 
$\pi B_{\nu}(T)$ in the limit $\kappa_{\rm {ff}}\gg \kappa_{\rm {T}}$ and becomes 
modified black body spectrum
\begin{equation}
F_{\nu}= 2\pi B_{\nu}(T)\sqrt{\kappa_{\rm {ff}}/\kappa_{\rm {T}}}\label{eqn4.8a}
\end{equation}
in the limit
$\kappa_{\rm {ff}}\ll \kappa_{\rm {T}}$. 
A part of the disc can emit black body spectrum
at lower frequencies $\nu\ll\nu_0$ and modified black body spectrum at
higher frequencies $\nu\gg \nu_0$, where $\nu_0=\nu_0(\rho,T)$ 
is defined such that $\kappa_{\rm {ff}}(\nu_0)=\kappa_{\rm {T}}$. 
However, if frequency 
averaged free-free opacity is larger than Thomson opacity,
${\bar \kappa}_{\rm {ff}}(T,\rho) > \kappa_{\rm {T}}$, then $h\nu_0>
kT$ and almost all the radiation is emitted as a black body spectrum. 
In the opposite limit, ${\bar \kappa}_{\rm {ff}}(T,\rho) \ll \kappa_{\rm {T}}$, 
one has $h\nu_0\ll kT$ and the spectrum is mostly modified black body,
transiting to black body only at lowest frequencies, $\nu<\nu_0$. 
Dotted line on Figs.~\ref{fig_m8d54}--\ref{fig_m1d54} separates 
regions with ${\bar \kappa}_{\rm {ff}}> \kappa_{\rm {T}}$ and 
${\bar \kappa}_{\rm {ff}} < \kappa_{\rm {T}}$.
We see that the optically thick disc has 
${\bar \kappa}_{\rm {ff}} \ll \kappa_{\rm {T}}$ and emits
a modified black body spectrum in the inner parts but may become absorption 
dominated in the cooler outer parts, where black body spectrum will be emitted.
At lower accretion rates, $l_{\rm {E}}/(2\epsilon)<10^{-4}$, all the surface of the 
disc will emit black body spectrum (with different $T$ at different radii, 
of course). The locations of the regions with mostly black body and modified 
black body spectra over the disc radii are fairly insensitive to the 
black hole mass $M$. The surface temperature $T$ in formula~(\ref{eqn4.8}) should be 
determined by equating the total emitted flux, the integral of $F_{\nu}$ over
all frequencies, to the half of the heat production rate $Q$ per unit disc surface 
(half is to account for two surfaces of the disc). Introducing variable $\displaystyle
x=\frac{h\nu}{kT}$ one can write this energy balance condition as
\begin{equation}
\frac{Q}{2}=4\pi \frac{k^4 T^4}{h^3 c^2}\int_0^{\infty}\frac{x^3 dx}{(e^x-1)
\left(1+\sqrt{1+\kappa_{\rm {T}}/\kappa_{\rm {ff}}}\right)}\label{eqn4.9}\mbox{.}
\end{equation}   
Further introducing Stefan--Boltzmann constant 
$\displaystyle \sigma=\frac{2\pi^5 k^4}{15 c^2 h^3}$ into right hand side 
of equation~(\ref{eqn4.9}) and effective temperature
$Q/2=\sigma T_{\rm {eff}}^4$ (equation~(\ref{eqn3.6})) into left hand side of 
equation~(\ref{eqn4.9}) we can transform equation~(\ref{eqn4.9}) to
\begin{equation}
\frac{T^4}{T_{\rm {eff}}^4}=\frac{\pi^4/15}{\displaystyle\int_0^{\infty}
\frac{2}{1+\sqrt{1+\kappa_{\rm {T}}/\kappa_{\rm {ff}}}}\frac{x^3 dx}{e^x-1}}\mbox{.}
\label{eqn4.10}
\end{equation}
Since $\displaystyle \frac{\pi^4}{15}=\int_0^{\infty}\frac{x^3 dx}{e^x-1}$,
$T$ is always larger than $T_{\rm {eff}}$. This is in accordance with general 
thermodynamic argument that the black body is the most efficient 
emitter of all. Equation~(\ref{eqn4.10}) together with 
expression~(\ref{eqn4.5}) for $\kappa_{\rm {ff}}$, expression~(\ref{eqn19}) for 
$\rho$, expression~(\ref{eqn3.6}) for $T_{\rm {eff}}$, 
and $\kappa_{\rm {T}}=0.4\,\mbox{cm}^2\mbox{g}^{-1}$ has been solved 
numerically to determine $T(r)$. After one knows $T=T(r)$,
it is possible to integrate $F_{\nu}(r)$~(\ref{eqn4.8})  
over the disc surface to obtain the spectral distribution of the total energy 
emitted by the disc
\begin{equation}
E_{\nu}=2\int_{r_{\rm {in}}}^{r_{\rm {out}}} 2\pi r F_{\nu}\,dr \label{eqn4.11}\mbox{,}
\end{equation}  
where the $2$ accounts for the two surfaces of the disc.

When a significant interval of radii exists where the emitted spectrum is a modified
black body, e.g. $h\nu_0<kT$, it is possible to get an approximate analytic 
expression for $E_{\nu}$. For $\nu>\nu_0$ we use expression~(\ref{eqn4.8a})
for $F_{\nu}$, which becomes
\begin{equation}
F_{\nu}=2.6\times 10^{-3}\,\frac{\mbox{erg}}{\mbox{s}\,\mbox{cm}^2\,\mbox{Hz}}
T^{5/4}\rho^{1/2}x^{3/2}\frac{e^{-x}}{\sqrt{1-e^{-x}}}\label{eqn4.12}\mbox{.}
\end{equation}
Integrating expression~(\ref{eqn4.12}) over 
frequencies by integrating from $0<x<\infty$, we obtain
\[
\frac{Q}{2}=8.2\times 10^7\,\frac{\mbox{erg}}{\mbox{s}\,\mbox{cm}^2} T^{9/4}
\rho^{1/2}\mbox{,}
\]
which can be solved for the temperature using expression~(\ref{eqn11a}) 
for $Q$. Let us denote the temperature found in this way by $T_{\rm {s}}$. The 
expression for $T_{\rm {s}}$ is
\begin{eqnarray}
&& T_{\rm {s}}=2.1\times 10^{5}\,\mbox{K}\,\left(
\frac{l_{\rm {E}}}{2\epsilon}\right)^
{8/9}M_8^{-8/9}\left(\frac{B_{10}}{10^4\,\mbox{G}}\right)^{-4/3}
\times \nonumber\\
&& \left(\frac{r}{10 r_{\rm {g}}}\right)^{-8/3+4\delta/3} {\mathcal{G}}^{8/9}
\label{eqn4.13}\mbox{.} 
\end{eqnarray}
Now we use $T_{\rm {s}}$ to substitute in the expression~(\ref{eqn4.12}) for $F_{\nu}$,
and then to expression~(\ref{eqn4.11}) to obtain $E_{\nu}$. It is convenient
to switch from the integration in $r$ to the integration in $x$ in 
equation~(\ref{eqn4.11}). We do so by writing $\displaystyle dx=-\frac{h\nu}
{kT_{\rm {s}}^2}\frac{dT_{\rm {s}}}{dr}dr$ and expressing $x$ through $r$ for a given $\nu$
from equation~(\ref{eqn4.13}). This procedure can be done analytically if 
one puts ${\mathcal{G}}=1$, that is our analytical approximation do not describe 
spectrum emerging close to
the inner edge of the disc, where ${\mathcal{G}}$ deviates from $1$ significantly.
Carrying out calculations one obtains
\begin{eqnarray}
&& E_{\nu}=4.2\times 10^{44}\times 10^{-\frac{11.73}{2-\delta}}\,
\frac{\mbox{erg}}{\mbox{s}\,\mbox{Hz}}\,
\left(\frac{l_{\rm {E}}}{2\epsilon}\right)^{\frac{4-3\delta}{3(2-\delta)}}
\times \nonumber\\
&& M_8^{\frac{8-3\delta}{3(2-\delta)}} \left(\frac{B_{10}}{10^4\,\mbox{G}}\right)^
{\frac{1}{2-\delta}}\frac{1}{(2-\delta)} 
\nu^{\frac{4\delta-5}{4(2-\delta)}} \times \label{eqn4.14} \\
&& \int_{x_{\rm {in}}}^{x_{\rm {out}}}\,x^{\frac{3(2\delta-3)}{4(\delta-2)}}
\frac{e^{-x}}{\sqrt{1-e^{-x}}}\,dx
\mbox{,} \nonumber
\end{eqnarray} 
where $\displaystyle x_{\rm {in}}=\frac{h\nu}{kT_{\rm {s}}(r_{\rm {in}})}$ and 
$\displaystyle x_{\rm {out}}=\frac{h\nu}{kT_{\rm {s}}(r_{\rm {out}})}$. 
We take for typical 
estimates $r_{\rm {in}}=10r_{\rm {g}}$ and $r_{\rm {out}}=1000 r_{\rm {g}}$ 
as the inner and outer 
edges of the disc. Although inner parts of the disc contribute significantly 
to the total emitted power and determine the most energetic part of the 
spectrum, the calculation of the spectrum there must be performed by 
taking into account factor ${\mathcal{G}}$, not to mention relativistic effects. 
The outer extension of the disc at $1000 r_{\rm {g}}$ is 
somewhat arbitrary, but the disc beyond $1000 r_{\rm {g}}$ is too cool to be 
described by our simple radiation model and, in the case of AGNs, even 
the existence of the disc for $r\geq 1000 r_{\rm {g}}$ is questionable because of
the instability to the gravitational fragmentation.
The value of the integral in formula~(\ref{eqn4.14}) decreases exponentially
for $x_{\rm {in}}>1$. This corresponds to an exponential cutoff in the spectrum
for $h\nu>kT(r_{\rm {in}})$. If $x_{\rm {in}}\ll 1$ 
but $x_{\rm {out}}\geq 1$, then the value 
of the integral is almost independent on $\nu$ and is a slowly varying 
function of $\delta$. We see that $\displaystyle E_{\nu}\propto 
\nu^{(4\delta-5)/(8-4\delta)}$ in this case. Thus, $E_{\nu}$ 
is rising for $\delta>5/4$ and declining for $\delta<5/4$. If both $x_{\rm {in}}
\ll 1$ and $x_{\rm {out}}\ll 1$, then it is possible to see from the expansion of 
the integral in expression~(\ref{eqn4.14}) that $E_{\nu}\propto \nu$. 
At frequencies below $h\nu=kT(r_{\rm {out}})$ the whole disc surface would contribute
with the low frequency tails of modified black body spectra, which scale 
as $F_{\nu}\propto \nu$  (equation~(\ref{eqn4.12})). 
Therefore, it is easy to understand the scaling $E_{\nu}\propto \nu$ for 
$h\nu< kT(r_{\rm {out}})$. However, we do not see the latter spectral index in 
calculated spectra because $\kappa_{\rm {ff}}$ becomes comparable to $\kappa_{\rm {T}}$ 
already at the frequencies where $x_{\rm {out}}>1$. The $\displaystyle E_{\nu}\propto 
\nu^{(4\delta-5)/(8-4\delta)}$ law extends down to the frequency 
at which $x_{\rm {in}}=x_{0\rm {s}}(r_{\rm {in}})$, where 
$\displaystyle x_{0\rm {s}}=\frac{h\nu_0}{kT_{\rm {s}}(r)}$ is
found by equating $\kappa_{\rm {ff}}=\kappa_{\rm {T}}$. 
For $x_{0\rm {s}}\ll 1$ one obtains using 
expression~(\ref{eqn4.5}) for $\kappa_{\rm {ff}}$ together with 
expression~(\ref{eqn4.13}) for $T_{\rm {s}}$ and expression~(\ref{eqn19}) for $\rho$
\begin{eqnarray}
&& x_{0\rm {s}}\approx 1.5\times 10^{-1}\left(\frac{l_{\rm {E}}}{2\epsilon}
\right)^{-46/27}
M_8^{46/27}\left(\frac{B_{10}}{10^4\,\mbox{G}}\right)^{32/9} 
\times \nonumber\\
&& \left(\frac{r}{10 r_{\rm {g}}}\right)^{46/9-32\delta/9}{\mathcal{G}}^{-46/27}
\label{eqn4.15}\mbox{.}
\end{eqnarray}
For $x<x_{0\rm {s}}$ spectrum gradually transits to the sum of a local black body 
with $T=T_{\rm {eff}}$, which has $E_{\nu}\propto \nu^{1/3}$
\citep{shakura73, shapiro83, krolik99}. Finally, for $h\nu \ll kT(r_{\rm {out}})$
the spectrum is the sum of $\propto \nu^2$ low frequency black bodies
of $\nu<\nu_0$. When the outer part of the disc, 
where ${\bar \kappa}_{\rm {ff}}>\kappa_{\rm {T}}$, is sufficiently truncated 
then the
$E_{\nu}\propto \nu^{1/3}$ part of the spectrum may be absent and the spectrum 
will transit directly from $\displaystyle E_{\nu}\propto \nu^{(4\delta-5)/(8-4\delta)}$
power to $E_{\nu}\propto \nu^2$ power. 

In summary, we see that magnetically dominated accretion discs 
have power law spectra with the spectral index depending on the radial 
distribution of magnetic field strength such that, 
$E_{\nu}\propto \nu^{(4\delta-5)/(8-4\delta)}$. 
This contrasts the standard weakly magnetized $\alpha$-disc which shows
a declining modified black body formed from the inner radiation dominated disc 
with $E_{\nu}\propto \nu^{-2/5}$. 

\subsection{Modified black body spectrum in a standard disc}
\label{subsec42}

As a side remark we note that the value for 
the spectral index $\displaystyle \gamma=\frac{\nu d\ln E_{\nu}}{d\nu}$ close 
to $0$ found for the latter regime of accretion disc by \citet{shakura73}
(text on page~349 after equation~[3.11] of that work) is different from ours 
$\gamma=-2/5$. It is easy to follow the exact prescription of \citet{shakura73},
namely calculate integral [3.10] in their work for spectrum [3.2] and temperature
[3.7]. As a result we obtain $\gamma=-2/5$ rather than $\gamma=0.04$ given in 
\citet{shakura73}. We need to point out this discrepancy because it is 
widely stated in many textbooks on accretion discs \citep{shapiro83, krolik99} 
with the reference to \citet{shakura73} that high luminosity accretion discs 
has almost flat plateau in its spectrum before the exponential cut off corresponding 
to $kT(r_{\rm {s}})$. However, the flat spectrum $E_{\nu}\propto \nu^{1/29}$ is 
produced by part~(b) of the standard disc model, where gas pressure dominates 
over radiation pressure. The spectral index $\gamma=1/29$ is close to the 
$\gamma=0.04$ given in \citet{shakura73} but the radial dependence of the surface
temperature in zone~(b) is $T_{\rm {s}}\propto r^{-29/30}$ rather than 
$T_{\rm {s}}\propto r^{-5/3}$ given by their formula~[3.7].
Thus, the standard $\alpha$-disc possessing both (b) 
and (a) zones should have spectrum steepening from plateau to $\propto \nu^{-2/5}$
and then exponentially cutting off at the temperature of the inner edge. 
Because the intervals of $r$, where approximate analytic expressions for 
emitted spectrum are valid, do not typically exceed one order of magnitude (the 
same is true for our disc model as well), one does not see ``pure'' extended
power laws when calculating spectra numerically by using general 
expressions~(\ref{eqn4.8}) and~(\ref{eqn4.9}). For example, \citet{wandel88} found 
$\gamma=-2/5$ slope in the narrow interval 
between 1000~\AA{} and 1450~\AA{} by numerically integrating disc spectra.

\subsection{Results of spectrum calculations}
\label{subsec43}

\begin{figure}
\includegraphics[width=\textwidth]{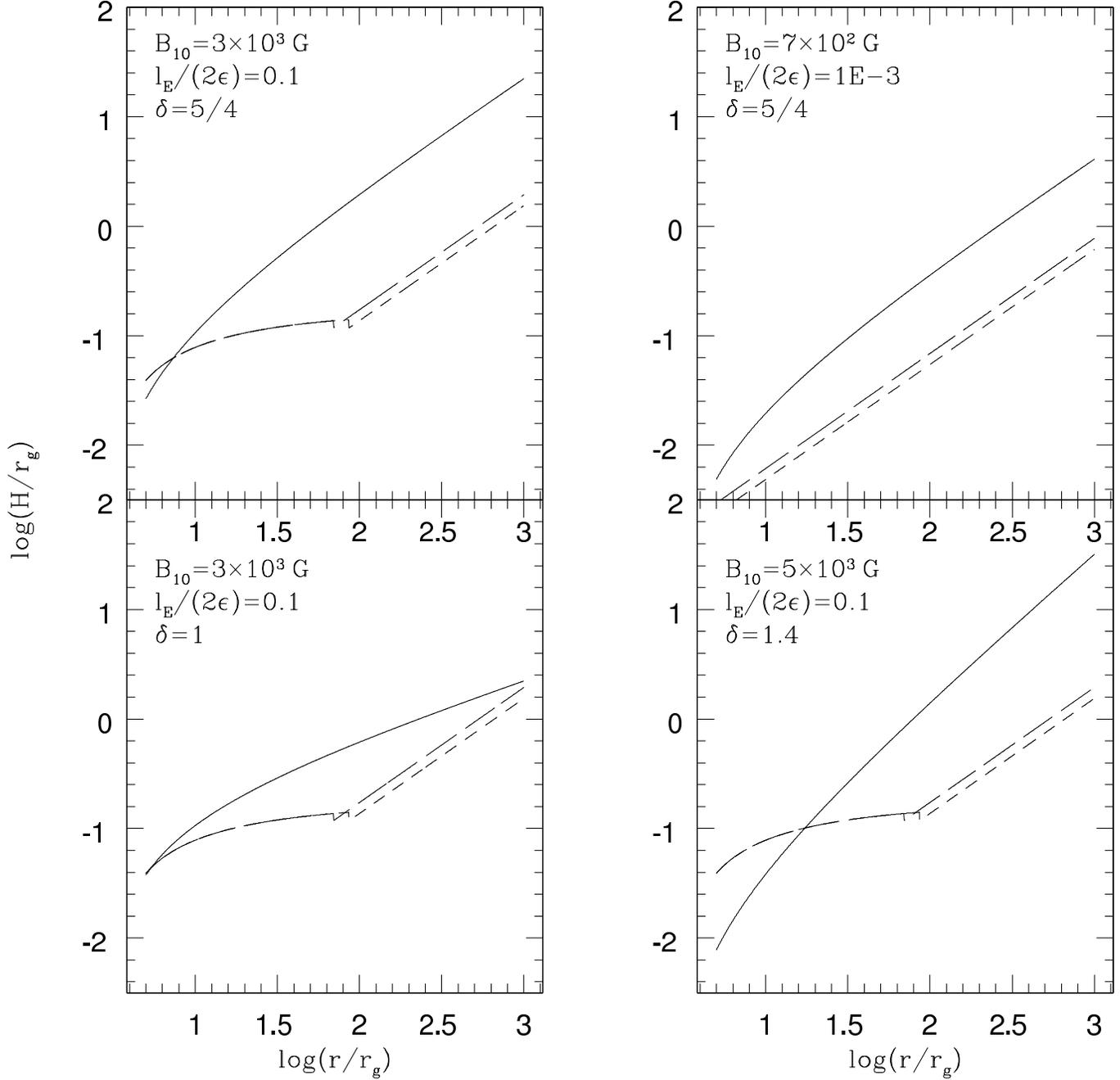}
\caption{Dependencies of the half-thickness of the disc $H$ on radius
for magnetically dominated disc (solid line) and Shakura--Sunyaev disc
with the same $l_{\rm {E}}/\epsilon$ and $M_8$ parameters and viscosity parameter
$\alpha=0.1$ (short-dashed line) and $\alpha=0.01$ (long-dashed line).
The breaks in the curves for the Shakura--Sunyaev disc occur at the interface
of zones~(a) and~(b) and are the results of using approximate analytic 
expressions in zone~(a) and zone~(b). For $l_{\rm {E}}/(2\epsilon)=10^{-3}$
zone~(b) extends down to the inner edge of the disc.
}
\label{fig_hh4}
\end{figure}

\begin{figure}
\includegraphics[width=\textwidth]{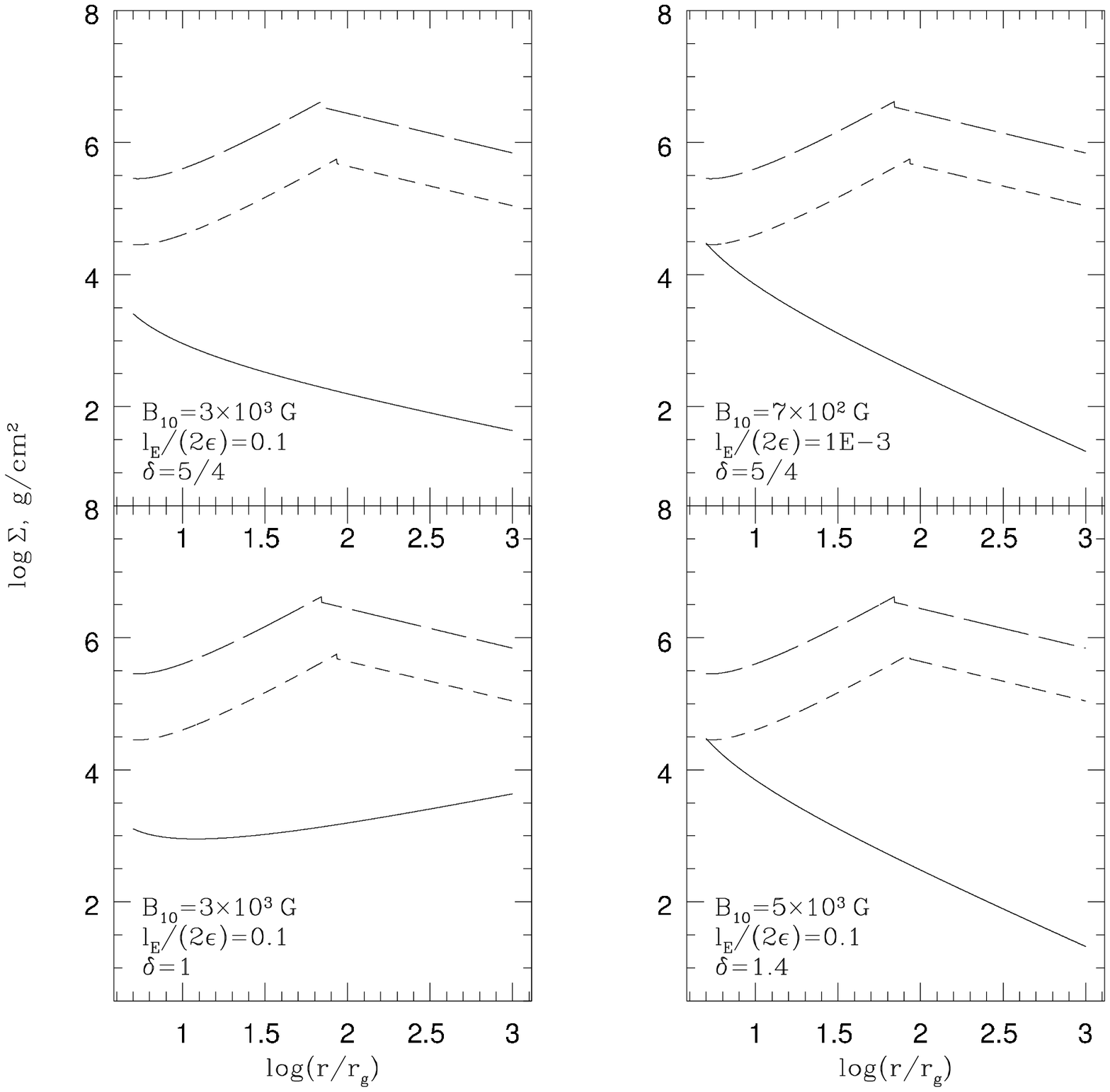}
\caption{Dependencies of the surface density $\Sigma$ on radius for 
magnetically dominated disc (solid line) and Shakura--Sunyaev disc
with the same $l_{\rm {E}}/\epsilon$ and $M_8$ parameters and viscosity parameter
$\alpha=0.1$ (short-dashed line) and $\alpha=0.01$ (long-dashed line).
The breaks in the curves for the Shakura--Sunyaev disc occur at the interface
of zones~(a) and~(b) and are the results of using approximate analytic 
expressions in zone~(a) and zone~(b).  For $l_{\rm {E}}/(2\epsilon)=10^{-3}$
zone~(b) extends down to the inner edge of the disc.
}
\label{fig_sigma4}
\end{figure}

\begin{figure}
\includegraphics[width=\textwidth]{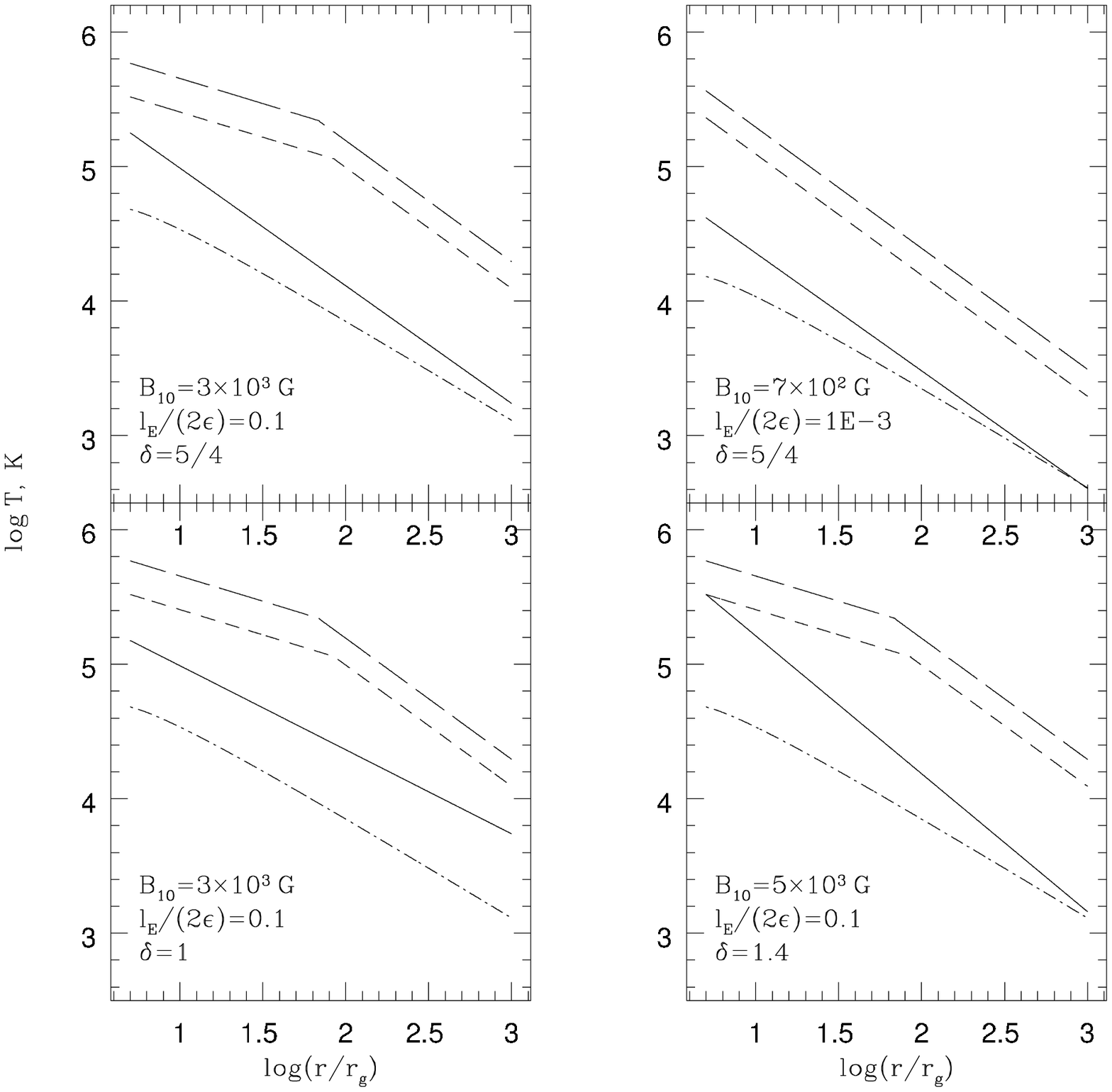}
\caption{Dependencies of temperatures on radius: $T_{\rm {mpd}}$ 
for magnetically dominated disc (solid line); $T_{\rm {mpd}}$ for 
Shakura--Sunyaev disc with the same $l_{\rm {E}}/\epsilon$ and $M_8$ parameters 
and viscosity parameter
$\alpha=0.1$ (short-dashed line) and $\alpha=0.01$ (long-dashed line);
$T_{\rm {eff}}$ (dashed-dotted line).
The breaks in the curves for the Shakura--Sunyaev disc occur at the interface
of zones~(a) and~(b) and are the results of using approximate analytic 
expressions in zone~(a) and zone~(b). For $l_{\rm {E}}/(2\epsilon)=10^{-3}$
zone~(b) extends down to the inner edge of the disc.
}
\label{fig_tmpd4}
\end{figure}

\begin{figure}
\includegraphics[width=\textwidth]{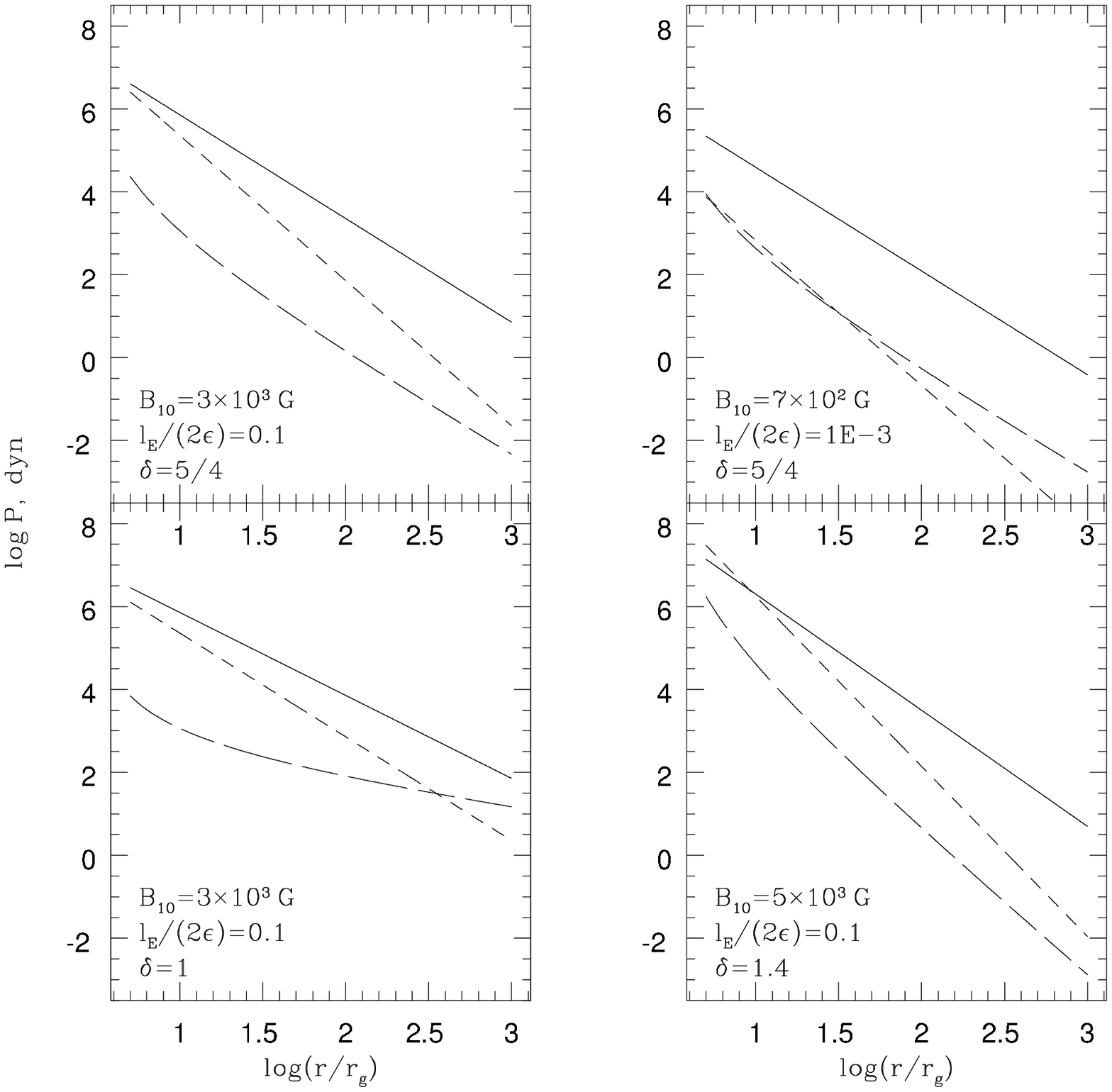}
\caption{Dependencies of pressures on radius for parameters of disc considered
in section~\protect\ref{sec4}. 
Magnetic plus turbulent pressure $B^2/(4\pi)$ is plotted with a solid line, 
$P_{\rm {rad}}$
is plotted with a short-dashed line, and $P_{\rm {th}}$ is plotted with a 
long-dashed line.}
\label{fig_press4}
\end{figure}

\begin{figure}
\includegraphics[width=\textwidth]{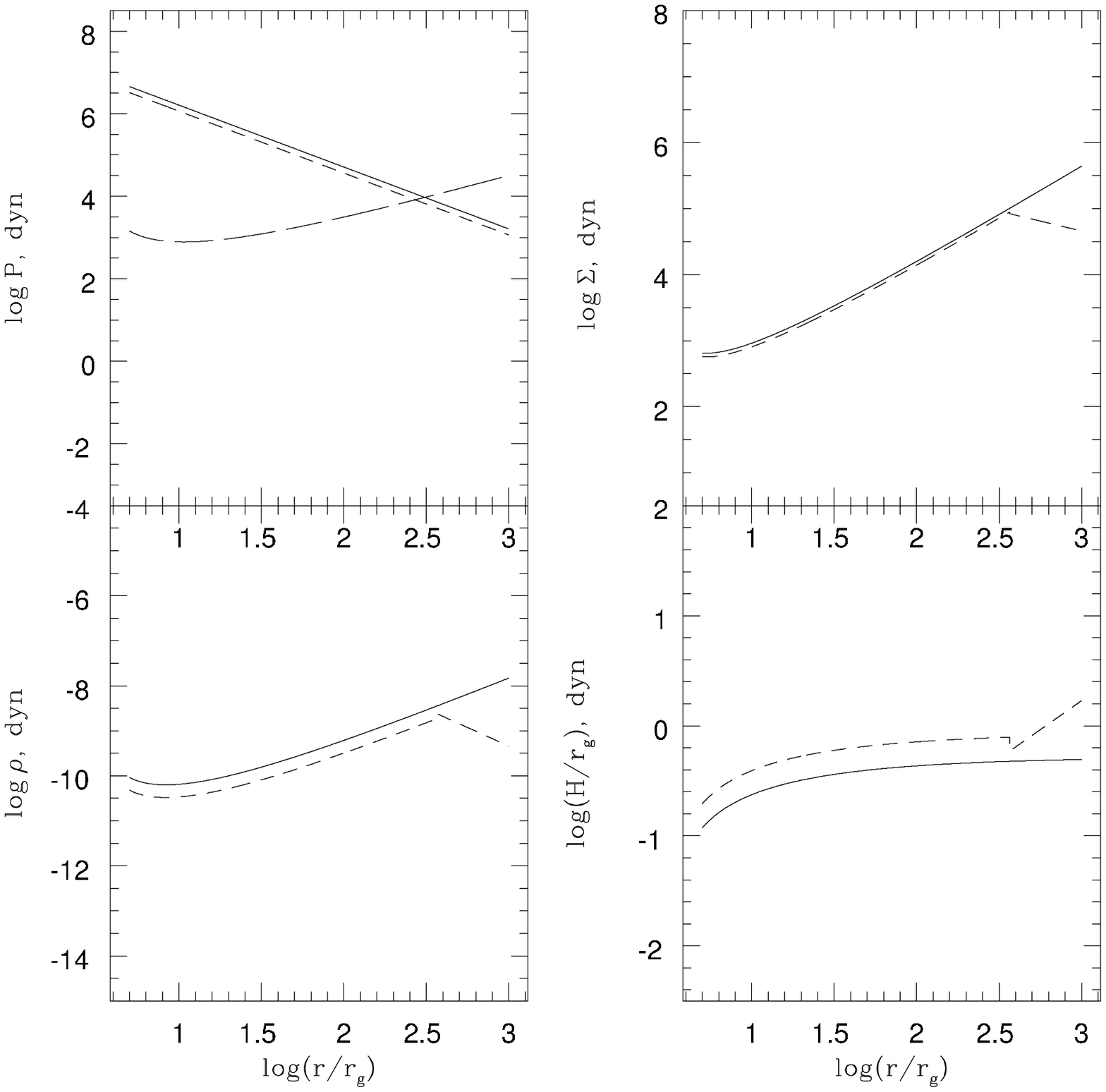}
\caption{Comparison of $\beta=1$ limit of magnetically dominated disc model 
with $\delta=3/4$, $l_{\rm {E}}/(2\epsilon)=0.5$, and $M_8=1$ and Shakura--Sunyaev
model. Top-left plot shows the radial profiles of magnetic plus turbulent 
pressure $=B^2/(4\pi)$ (solid line), $P_{\rm {rad}}$ (short-dashed line), 
and $P_{\rm {th}}$ (long-dashed line) in the magnetically dominated model. 
Bottom-left, top-right,
and bottom-right plots show the radial profiles of $\rho$, $\Sigma$, and $H$
in the magnetically dominated model (solid lines) and the 
Shakura--Sunyaev model with $\alpha=1$ (dashed lines).  
The parameters $l_{\rm {E}}/(2\epsilon)$ and 
$M_8$ are the same for magnetically dominated 
model and Shakura--Sunyaev model.
}
\label{fig_beta1}
\end{figure}

\begin{figure}
\includegraphics[width=\textwidth]{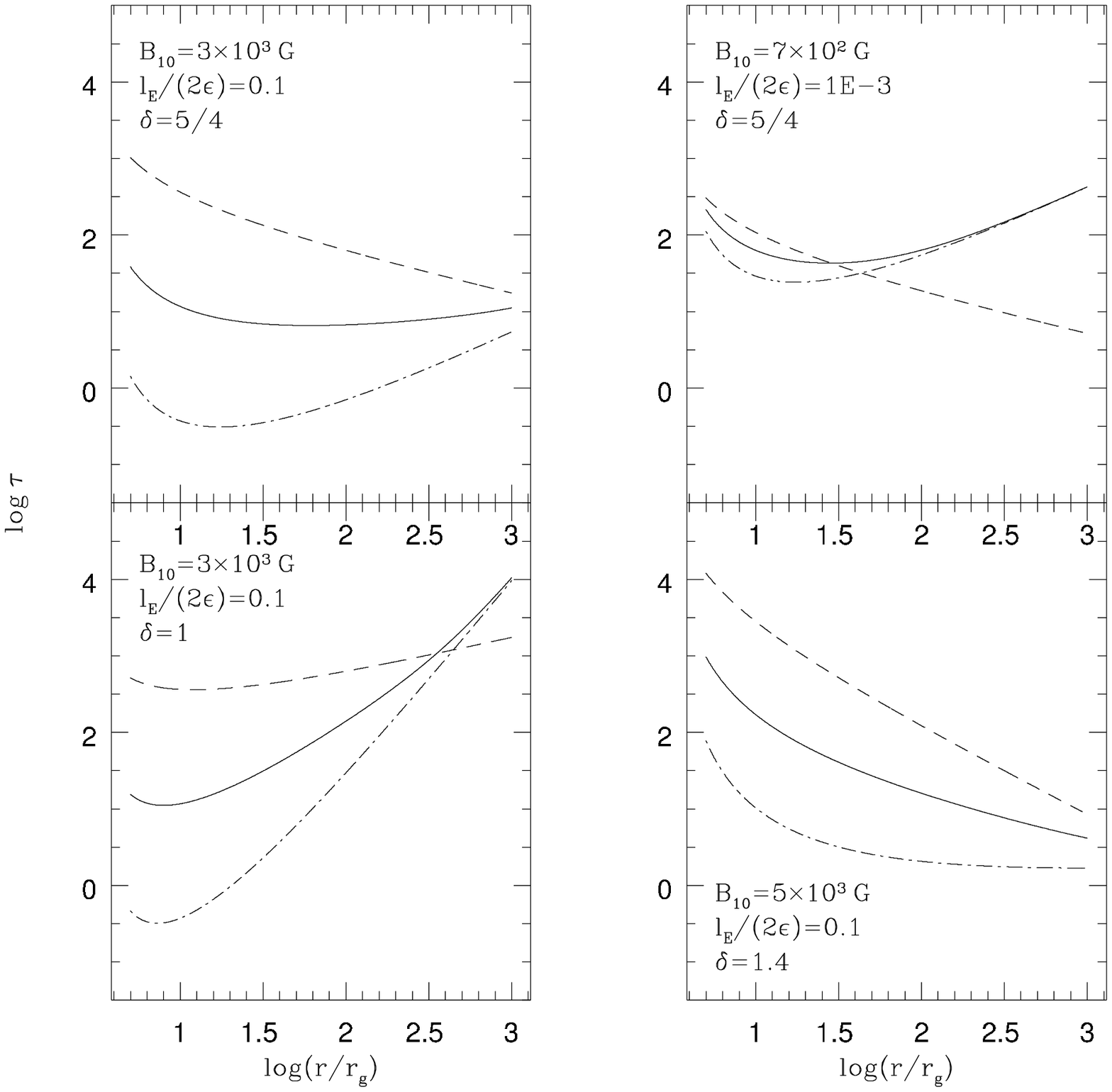}
\caption{Dependencies of optical depth through the half disc thickness 
on radius for parameters of disc considered in 
section~\protect\ref{sec4}. Dashed line is $\tau_{\rm {c}}$, dashed-dotted line is 
${\bar\tau}_{\rm {ff}}$, and solid line is ${\bar\tau}_{\ast}$.
}
\label{fig_opac4}
\end{figure}

\begin{figure}
\includegraphics[width=\textwidth]{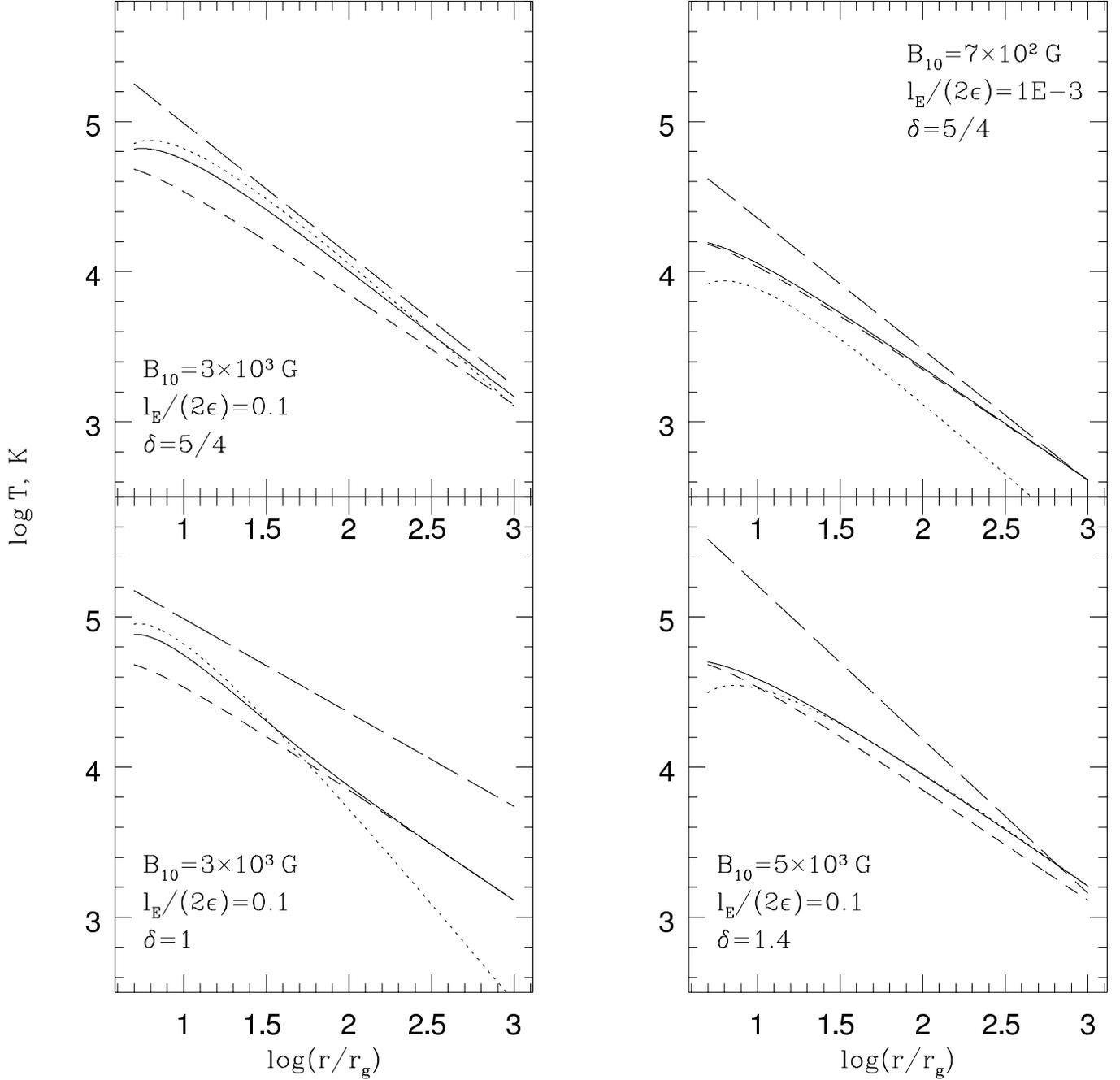}
\caption{Dependencies of temperatures on radius for parameters of disc considered in 
section~\protect\ref{sec4}. Short dashed line is $T_{\rm {eff}}$, long dashed line is
$T_{\rm {mpd}}$, dotted line is $T_{\rm {s}}$, and solid line is $T$. The latter two temperatures 
are the surface temperatures of the disc calculated in section~\ref{sec4}.
}
\label{fig_temp4}
\end{figure}

\begin{figure}
\includegraphics[width=\textwidth]{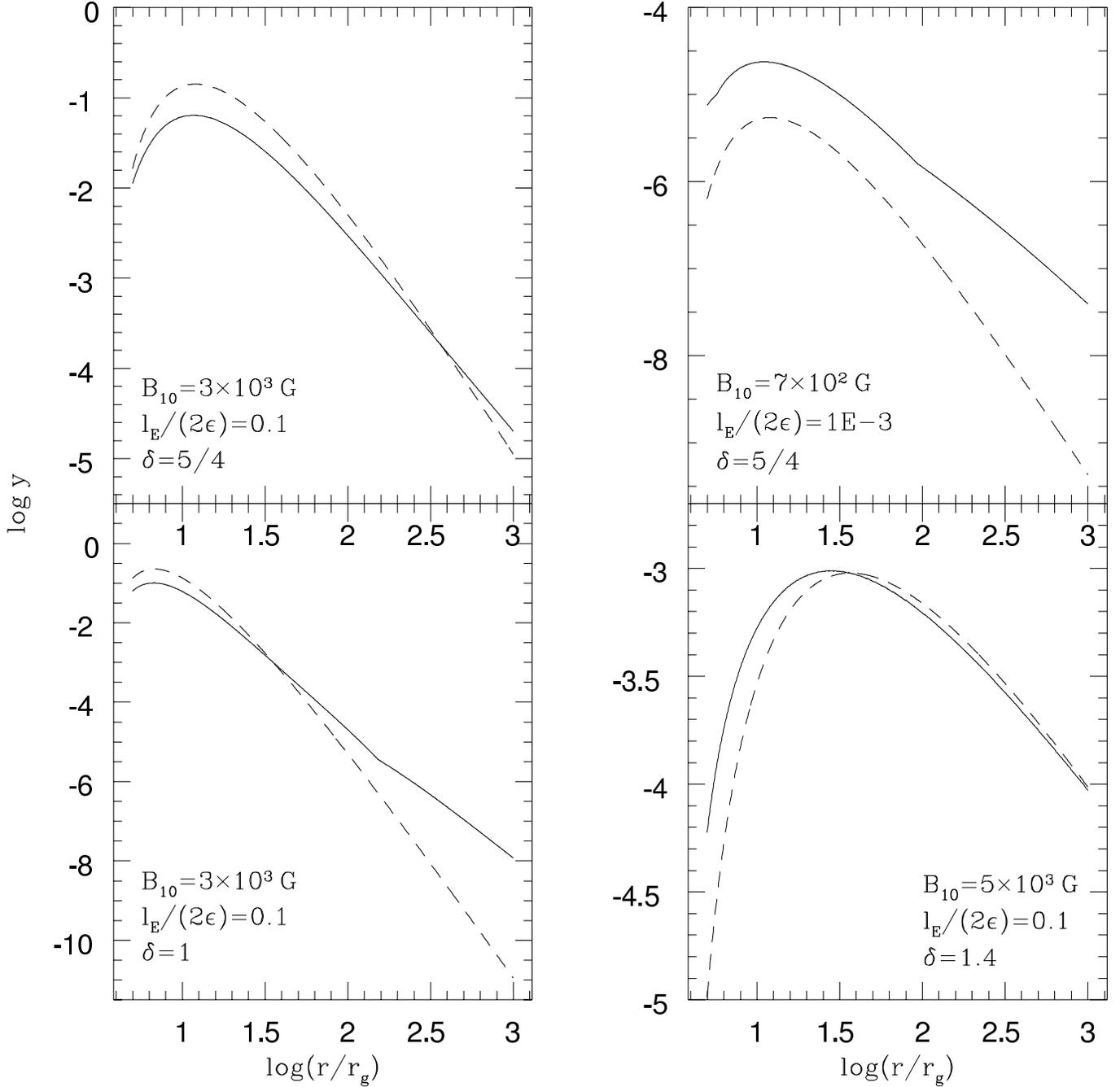}
\caption{Dependencies of Comptonisation parameter $y$ on radius for parameters 
of disc considered in section~\protect\ref{sec4}. Solid line is for $y$ 
calculated using temperature~$T$, exact expression~(\protect\ref{eqn4.7}) 
for $\tau_{\rm {es}}$, and $h\nu=5kT(r)$. 
Dashed line is for $y$ calculated using $T_{\rm {s}}$, assuming 
that $\tau_{\rm {es}}\approx (\tau_{\rm {T}}/\tau_{\rm {ff}})^{1/2}$, 
and $h\nu=5kT_{\rm {s}}$.
}
\label{fig_y4}
\end{figure}

\begin{figure}
\includegraphics[width=\textwidth]{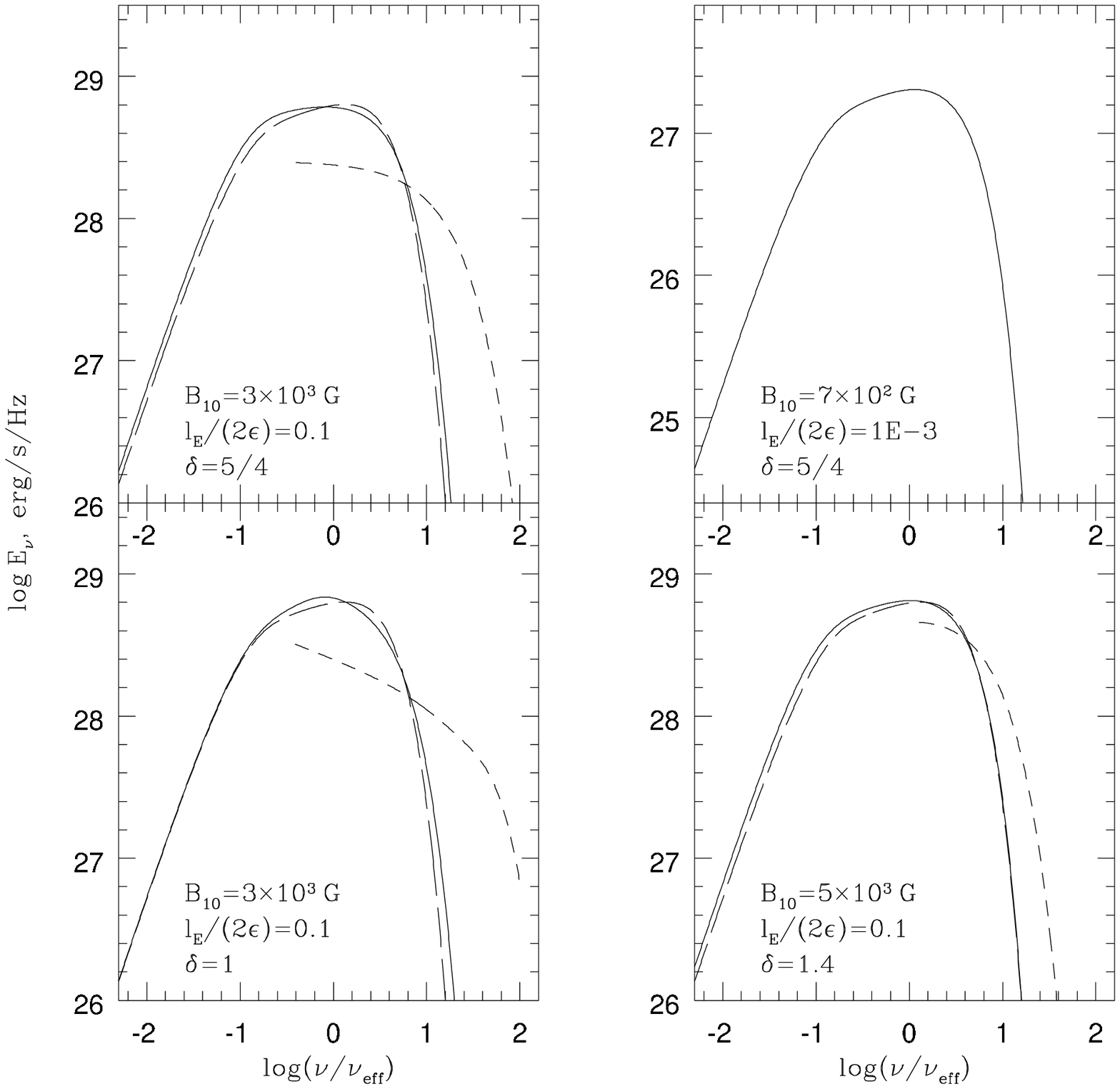}
\caption{Spectral energy distribution for the total flux from the disc. Frequency
is plotted in units of $\nu_{\rm {eff}}=kT_{\rm {eff}}(10r_{\rm {g}})/h$. 
Exact values of $E_{\nu}$ 
calculated using temperature $T$ are plotted with solid lines; values of $E_{\nu}$
calculated from analytical expression~(\protect\ref{eqn4.14}) are plotted with 
short-dashed lines. The latter are shown only for frequencies at which $x_{\rm {in}}$ is 
larger than the minimal value of $x_{0\rm {s}}(r)$, 
which is achieved at about $5r_{\rm {g}}$ to $10r_{\rm {g}}$. 
Spectra of Shakura--Sunyaev discs are plotted for viscosity parameter $\alpha=0.1$ with 
long-dashed lines.  
}
\label{fig_spectr4}
\end{figure}

We present results of the simplified analytical integration of the spectrum
using equation~(\ref{eqn4.12}) as well as more exact numerical integration 
using equation~(\ref{eqn4.8}), solving for $T$ from equation~(\ref{eqn4.10}) 
and integrating equation~(\ref{eqn4.11}). 
Function ${\mathcal{G}}(r)$ was kept
in numerical calculations, so the results are applicable to the innermost 
parts of the disc, where the most of energy is radiated.
For a given $M_8$ and $l_{\rm {E}}/\epsilon$,
an optically thick magnetically dominated discs exist within 
$5r_{\rm {g}}<r<1000 r_{\rm {g}}$   
only for $\delta$ in the interval of about $1$ to $1.4$. In Figs.
\ref{fig_hh4}--\ref{fig_spectr4} 
we illustrate models
for the four choices of parameter sets:  
\begin{enumerate}
\item $M_8=1$, $\frac{l_{\rm {E}}}{2\epsilon}=0.1$,
$\delta=5/4$, $B_{10}=3\times 10^3\,\mbox{G}$, 
\label{parameter1}
\item $M_8=1$, $\frac{l_{\rm {E}}}{2\epsilon}=0.1$, $\delta=1$, 
$B_{10}=3\times 10^3\,\mbox{G}$,
\label{parameter2}
\item $M_8=1$, $\frac{l_{\rm {E}}}{2\epsilon}=0.1$, $\delta=1.4$, 
$B_{10}=5\times 10^3\,\mbox{G}$, 
\label{parameter3}
\item $M_8=1$, $\frac{l_{\rm {E}}}{2\epsilon}=10^{-3}$,
$\delta=5/4$, $B_{10}=7\times 10^{2}\,\mbox{G}$. 
\label{parameter4}
\end{enumerate}

The dependencies of $H$ on $r$ given by equation~(\ref{eqn17}) for four
parameter sets are plotted in Fig.~\ref{fig_hh4}. For comparison we also plot 
the half-thickness $H$ in the Shakura--Sunyaev model of the disc with 
the same accretion rate (parametrized by $l_{\rm {E}}/\epsilon$) and the same 
mass of the central object. We use approximate analytic expressions 
for the parameters of the disc ($H$, $\rho$, $\Sigma$, $T_{\rm {mpd}}$) 
in the radiation dominated zone~(a) of the Shakura--Sunyaev model 
and thermal pressure dominated zone~(b) \citep{shakura73}. 
The magnetically dominated disc is thicker than the 
standard disc. For higher accretion rates, the standard disc has a concave 
shape due to the transition from inner zone~(a) to intermediate zone~(b),
which allows illumination of the outer parts of the disc by the inner
hotter and brighter parts.
Magnetically dominated disc has convex shape, which exclude such 
illumination.

In Fig.~\ref{fig_sigma4} we plot the dependencies of 
the column thickness through the disc $\Sigma$ (equation~(\ref{eqn18})) 
on the radius and also 
compare to Shakura--Sunyaev standard disc. The magnetically dominated disc 
is much less massive than the standard disc. Both $\Sigma$ and 
$\rho$ are smaller for magnetically dominated discs, and only in the inner 
$\sim 10 r_{\rm {g}}$ are the densities comparable.

The dependencies of mid-plane temperature $T_{\rm {mpd}}$ on radius 
given by equation~(\ref{eqn3.8}) are shown in Fig.~\ref{fig_tmpd4}. On 
the same figure we also plot $T_{\rm {mpd}}$ in the Shakura--Sunyaev model
and $T_{\rm {eff}}$ given by equation~(\ref{eqn3.6}), which is the same 
for magnetically dominated and standard discs. Because of the lower column
density of the magnetically dominated disc, $T_{\rm {mpd}}$ is less 
than for standard $\alpha$-discs.

The dependencies of magnetic plus turbulent pressure, $\simeq B^2/(4\pi)$, 
radiation pressure in the disc mid-plane $P_{\rm {rad}}=a T_{\rm {mpd}}^4/3$, and 
thermal pressure in the disc mid-plane $P_{\rm {th}}=nkT_{\rm {mpd}}$
are presented in Fig.~\ref{fig_press4}. We see that the assumption of 
magnetic pressure dominance is well satisfied for our models except in the 
innermost regions, $r\la 10 r_{\rm {g}}$, for higher accretion rates 
$l_{\rm {E}}/(2\epsilon)=0.1$, where radiation pressure becomes comparable to 
the magnetic pressure. The latter fact limits the existence of magnetically
dominated regime in the innermost parts of accretion discs for higher 
luminosities. Plasma parameter $\beta$ 
defined as $\beta=8\pi (P_{\rm {rad}}+P_{\rm {th}})/B^2$ decreases
with radius and varies from $\sim 1$ to $\sim 10^{-2}$ in our models.
 
In the limit $\beta=1$ magnetic pressure is comparable to the largest of radiation
or thermal pressures and our strongly magnetized disc model transforms into 
Shakura--Sunyaev model with $\alpha=1$. If $\delta=3/4$ in our model, 
then the radial scaling of the magnetic and turbulent pressures, 
$B^2/(4\pi)$, is the same as that of the radiation pressure inside the disc, 
$aT^4_{\rm {mpd}}$, aside from the factor $\mathcal{G}$.
If $\delta=51/40$, 
then the radial scaling of 
$B^2/(4\pi)$ is the same as that of the thermal pressure inside the disc, 
$nkT_{\rm {mpd}}$. Therefore, by choosing 
$\delta=3/4$ and adjusting the magnitude of $B_{10}$ one can construct 
the model with approximately constant $\beta$ in the zone where radiation 
pressure exceeds thermal pressure. By choosing $\delta=51/40$ one can 
construct constant $\beta$ model in the zone where thermal pressure dominates
radiation pressure. 
We illustrate this in Fig.~\ref{fig_beta1},
where we show the dependencies of pressures, $\rho$, $H$, and $\Sigma$ on 
$r$ for our model with $\delta=3/4$, $l_{\rm {E}}/(2\epsilon)=0.5$, and $M_8=1$, 
and for the Shakura--Sunyaev model with $\alpha=1$ for the same accretion 
rate $l_{\rm {E}}/(2\epsilon)$ and $M_8$. The transition from zone~(a) to zone~(b) 
in this Shakura--Sunyaev model occurs at $r_{\rm {ab}}=360 r_{\rm {g}}$. 
The breaks on
the curves corresponding to the Shakura--Sunyaev model occur at $r=r_{\rm {ab}}$ 
in Fig~\ref{fig_beta1}. We adjusted $B_{10}$ such that the magnetic pressure 
will be in equipartition with the radiation pressure in our model. Then, as
it is seen from the top-left plot in Fig.~\ref{fig_beta1}, thermal pressure 
is less than magnetic and radiation pressures for $r$ less than some $r_{\rm {c}}$ 
and exceeds magnetic and radiation pressures for $r>r_{\rm {c}}$, so our model is not 
applicable for $r>r_{\rm {c}}$. The subsequent three plots show 
that $r_{\rm {c}}\approx r_{\rm {ab}}$. Two right plots and bottom-left 
plot in Fig.~\ref{fig_beta1}
show that $\rho$, $H$, and $\Sigma$ in our model for
$r<r_{\rm {c}}$ are very close to $\rho$, $H$, and $\Sigma$ in 
Shakura--Sunyaev $\alpha=1$ model in the radiation pressure 
dominated zone $r<r_{\rm {ab}}$. A
similar conclusion holds for the transition of our model with $\delta=51/40$ 
to a Shakura--Sunyaev zone~(b) model for $r>r_{\rm {ab}}$ and $\beta=1$. 
The radiation spectra of our model in the limit $\beta=1$ also approach that of the 
Shakura--Sunyaev model, as shown by direct numerical calculations. 
The power law modified black body spectrum 
$E_{\nu}\propto \nu^{(4\delta-5)/(8-4\delta)}$ derived in section~\ref{subsec41}
becomes $E_{\nu}\propto \nu^{-2/5}$ for $\delta=3/4$ and $E_{\nu}\propto \nu^{1/29}$
for $\delta=51/40$, which is coincident with the modified black body power laws 
for the Shakura--Sunyaev zone~(a) and zone~(b) spectra (see section~\ref{subsec42}).

The dependencies of optical depths through the half disc 
thickness on radius are shown in
Fig.~\ref{fig_opac4} for four parameter sets. The three curves plotted are: 
${\bar\tau}_{\rm {ff}}$ given by equation~(\ref{eqn3.15}), 
$\tau_{\rm {c}}=\kappa_{\rm {T}}\Sigma/2$, and
the effective optical thickness $\displaystyle {\bar \tau}_{\ast}=
\frac{\Sigma}{2}\sqrt{(\kappa_{\rm {T}}+
{\bar \kappa}_{\rm {ff}}){\bar \kappa}_{\rm {ff}}}$.
For $\delta=5/4$, the effective optical thickness $\displaystyle {\bar \tau}_{\ast}$
is almost constant throughout the disc, but when $\delta$ deviates from $5/4$, 
$\displaystyle {\bar \tau}_{\ast}$ starts to approach $1$ either at the inner or
at the outer edge of the disc and so our model breaks down at those radii.
With the decrease of the accretion rate, the disc becomes cooler and denser so the 
absorbing opacity rises and becomes larger than the scattering opacity in the 
outer parts of the disc. 

In Fig.~\ref{fig_temp4} we show: $T_{\rm {eff}}(r)$ given by equation~(\ref{eqn3.6}), 
$T_{\rm {mpd}}(r)$ given by equation~(\ref{eqn3.8}), $T_{\rm {s}}(r)$ 
given by equation~(\ref{eqn4.13}),
and $T(r)$ by solving equation~(\ref{eqn4.10}) numerically.
Note that $T_{\rm {eff}}$ and surface temperature $T$ are 
always smaller than the $T_{\rm {mpd}}$ for an optically thick disc. For low accretion rate, 
$l_{\rm {E}}/(2\epsilon)=10^{-3}$, ${\bar \kappa}_{\rm {ff}}>\kappa_{\rm {T}}$  
and $T\approx T_{\rm {eff}}$. In this case, $T_{\rm {s}}$ is ill defined and 
values of $T_{\rm {s}} < T_{\rm {eff}}$ are unphysical on the plot for
$l_{\rm {E}}/(2\epsilon)=10^{-3}$ and also for $r>100 r_{\rm {g}}$ on the plot for 
$\delta=1$, $l_{\rm {E}}/(2\epsilon)=0.1$. The temperature $T_{\rm {s}}$ becomes
a good approximation for $T$ when $T_{\rm {s}}>T_{\rm {eff}}$ (scattering dominates 
over absorption in the surface layer). Unlike the values and slope of $T_{\rm {mpd}}(r)$, 
which substantially increases with increasing $\delta$, the value of $T$ is less 
sensitive to $\delta$: only inner parts of the disc becomes slightly 
hotter for larger values of $\delta$. Both $T$ and $T_{\rm {eff}}$ are changed significantly 
when the accretion rate or mass $M$ are changed. 

Fig.~\ref{fig_y4} shows the 
results of calculating Comptonisation $y$ parameter according to 
equations~(\ref{eqn4.6}) and~(\ref{eqn4.7}). We conservatively set $x=5$ for 
the calculation of $y$. Then, $y$ is the function of radius alone. 
On the same figure we also show $y$ in the 
regime of modified black body spectrum, using $T_{\rm {s}}$ and writing the simplified 
version of equation~(\ref{eqn4.6}) as $\displaystyle 
y=\frac{4kT_{\rm {s}}}{m_{\rm {e}} c^2}
\frac{\kappa_{\rm {T}}}{\kappa_{\rm {ff}}}$. We see that Comptonisation is not 
important for our models even in the inner disc.

Energy spectra $E_{\nu}$ are presented in Fig.~\ref{fig_spectr4}.
We normalized frequency to the characteristic frequency of an effective 
black body from the inner disc, namely, we plot
versus $\nu/\nu_{\rm {eff}}$, where 
\[
h\nu_{\rm {eff}}=kT_{\rm {eff}}(10 r_{\rm {g}})=
6\,\mbox{eV}\times\left(\frac{l_{\rm {E}}}{2\epsilon}\right)^{1/4}
M_8^{-1/4}
\mbox{,} 
\]
so the spectra plotted cover the range from {\it EUV} to infrared 
for $M=10^8\,{\rm M}_{\sun}$. 
We checked that the 
total thermal energy emitted from the disc between $r_{\rm {in}}$ and $r_{\rm {out}}$ 
calculated as an integral of the spectrum over frequencies is equal to the
surface integral of the dissipation $Q$ (expression~(\ref{eqn11a})):
\begin{eqnarray}
&& E=\int_0^{\infty}E_{\nu}\,d\nu=\int_{r_{\rm {in}}}^{r_{\rm {out}}}2\pi r Q\,dr = 
\nonumber\\
&& 9.5\times 10^{45}\,\mbox{erg s}^{-1}\,\frac{l_{\rm {E}}}{\epsilon}M_8
\frac{r_{\rm {g}}}{r_{\rm {in}}} \times \label{eqn4.16} \\
&& \left[1-
\frac{r_{\rm {in}}}{r_{\rm {out}}}-\frac{2}{3}\zeta
\left(\sqrt{\frac{r_{\rm {s}}}{r_{\rm {in}}}}-\frac{r_{\rm {in}}}{r_{\rm {out}}}
\sqrt{\frac{r_{\rm {s}}}{r_{\rm {out}}}}\right)\right]
\mbox{.} \nonumber
\end{eqnarray}
All spectra shown in Fig.~\ref{fig_spectr4} were computed by integrating from 
$r_{\rm {in}}=3.1 r_{\rm {g}}$ to $r_{\rm {out}}=1000 r_{\rm {g}}$ 
with the factor ${\mathcal{G}}(r)$ taken into account.
We also show the spectrum calculated by using the approximate analytic 
expression~(\ref{eqn4.14}) for a modified black body in its regime of 
validity ($x_{\rm {in}}>x_{0\rm {s}}(r_{\rm {in}})$). 
Because the analytic expression was obtained by setting ${\cal G}=1$
in equation~(\ref{eqn4.13}) for $T_{\rm {s}}$, it overestimates 
the temperature in the inner parts of the disc by a factor of a few
and does not describe the high energy part of the spectra correctly.
The lower frequency at
which the sum of modified black bodies is still a good approximation, 
increases with 
the overall increase of absorption in the disc. For parameter set~\ref{parameter4} 
above, with 
$l_{\rm {E}}/(2\epsilon)=10^{-3}$, pure modified black body cannot be 
used at all, so the corresponding panel in Fig.~\ref{fig_spectr4} does not 
show a second curve. The spectrum for $\delta=5/4$ (solid line) shows flat
plateau extending by more than an order of magnitude in frequency in accordance
with analytical result. Although the declining top part
of the spectrum for $\delta=1$ and rising top part for $\delta=1.4$ are
apparent, the interval of frequencies, where a modified black body approximation works, 
becomes small and blends with the $\propto \nu^{1/3}$ spectrum of the sum of 
local black bodies. Thus no dependence on $\delta$ is evident here. 
The top right plot for low luminosity $l_{\rm {E}}/(2\epsilon)=10^{-3}$
is indistinguishable from the sum of the local black body spectra. 
All spectra behave like $\propto \nu^2$ for low frequencies. In Fig.~\ref{fig_spectr4} 
we also show spectra of a Shakura--Sunyaev $\alpha$-disc with the same 
$l_{\rm {E}}/\epsilon$ and $M_8$ parameters to compare with plots of a magnetically 
dominated disc with viscosity parameter $\alpha=0.1$. These spectra were 
calculated in the same way we calculated the spectra of the magnetically dominated disc:
first we found the surface temperature $T(r)$ by solving equation~(\ref{eqn4.10})  
with the $\rho(r)$ profile taken from standard $\alpha$-disc model, and then
integrated equation~(\ref{eqn4.11}) with $F_{\nu}$ given by 
expression~(\ref{eqn4.8}). For low accretion rates 
of order of $10^{-2}$ of Eddington accretion rate and smaller, 
the spectra of Shakura--Sunyaev disc are very close to the sum
of local black bodies with temperatures $T_{\rm {eff}}(r)$ 
\citep{wandel88}. In general the spectra of our 
magnetically dominated discs are close to the 
spectra of Shakura--Sunyaev discs, so it seems to be difficult to distinguish 
between them observationally. This means that sources with discs 
previously thought to be thermally supported could actually be magnetically
supported.

\section{Discussion and Conclusions}

We have found self-consistent solutions for thin, magnetically
supported turbulent accretion discs assuming the 
tangential stress $\displaystyle f_{\phi}=\alpha(r) \frac{B^2}{4\pi}$.
When compared to the standard $\alpha$-disc 
models \citep{shakura73} magnetically dominated discs have lower surface
and volume densities at the same accretion rate. This is due to the more 
efficient angular momentum transport by supersonic turbulence and strong
magnetic fields than the subsonic thermal turbulence of 
the standard model. As a result, magnetically dominated discs are lighter 
and are not subject to self-gravity instability. 
In the limit of plasma $\beta=1$, magnetic pressure is comparable to the 
largest of radiation or thermal pressures and our strongly magnetized disc 
model transforms into the Shakura--Sunyaev model with $\alpha=1$.

When we derived the disc structure, we made no explicit distinction 
between turbulent and magnetic pressure support and angular momentum transfer. 
As such, our model would be valid in any situation in which the magnetic
and turbulent kinetic energies are comparable to, or greater than
the thermal energy density. The assumption that the kinetic and magnetic
energies are nearly comparable is natural because 
turbulence should result in the amplification of 
small scale magnetic fields in highly conducting medium due to dynamo action.
Typically, in a sheared system, the magnetic energy can be even slightly
larger than turbulent kinetic energy since the magnetic energy gains
from the additional shear. We find that the thermal spectrum
from the surface of the magnetically dominated disc in the 
optically thick regime is close to the spectrum of the standard 
Shakura--Sunyaev disc.

The issue  arises as to how the magnetic field could reach sonic
or supersonic energy densities. To obtain sonic turbulence and 
produce a $\beta=1$ disc, the MRI
might be sufficient. To obtain a $\beta < 1$ supersonic turbulence
may require something else. One possibility in AGN appeals to the 
high density of stars in the central stellar cluster surrounding 
AGN accretion  discs. Passages of stars through the disc might be an 
external source of supersonic turbulence
analogous to the supernovae explosions being the source of supersonic 
turbulence in the Galaxy. 
Stars pass through the disc with the velocities of order
of Keplerian velocity, which is much larger than the sound speed in the disc. 
We consider the support of turbulence by star-disc collisions 
in Appendix~B and 
find that statistically speaking, star-disc collisions are unlikely to provide
enough energy to sustain supersonic turbulence in most AGN accretion discs,
however the possibility remains that a small number out of a large population
could become magnetically dominated.

Indeed whether a disc could ever really attain a magnetically dominated
state is important to understand. The  present 
answer from simulations is not encouraging, 
but not completely ruled out. Further global 
MHD simulations of turbulence in vertically stratified accretion discs with 
realistic physical boundary conditions are needed along
with more interpretation and analysis.
Magnetic helicity conservation for example,
has not been fully analyzed in global accretion disc simulations 
to date, and yet the large scale magnetic helicity can act as a sink
for magnetic energy since magnetic helicity inverse cascades.

As an intermediate step in assessing the viability of 
low $\beta$ discs,  it may be interesting to assess whether they
are stable. One can take, as an initial 
condition, the stationary 
model of the magnetically dominated accretion disc given by 
expressions~(\ref{eqn14}), (\ref{eqn17}--\ref{eqn19}) with the initial 
magnetic field satisfying all constraints of our model and falling into the 
shadowed regions on plots in Figs.~\ref{fig_m8d54}--\ref{fig_m1d54}.

One point of note is that 
magnetically dominated discs may be helpful (though perhaps
not essential, if large scale magnetic fields can be produced 
\citep{blackman02,
blackman03}) in explaining AGN sources in which 40\% of the bolometric
luminosity comes from hot X-ray coronae. 
If the non-thermal component in galactic black hole sources is 
attributed to the magnetized corona above the disc (e.g., \citet{dimatteo99},
also \citet{beloborodov99} discusses possible alternatives), then 
magnetically dominated discs can naturally explain large fractions, 
up to 80~\% \citep{dimatteo99}, of the accretion power being transported
into coronae by magnetic field buoyancy (although $\beta\la 1$ disc 
solutions are also possible, \citep{merloni03}).
Though coronae can form in systems with high $\beta$ interiors, 
the percentage of the dissipation
that goes on in the interior vs. the coronae could be $\beta$ dependent.

The main purpose of our study was simply to explore the consequences
of making a magnetically dominated analogy to Shakura and Sunyaev,
and filling in the parmeter regime which they did not consider.
In the same way that we cannot provide proof that a disc
can be magnetically dominated, 
they did not present proof that a disc must be turbulent, but
investigated the  consequences of their assumption.
We also realize that the naive $\alpha$ disc formalism itself can be questioned
and its ultimate validity in capturing
the real physics is limited. 
Nevertheless it still has an appeal of simplicity.

Finally, we emphasize that 
our model does not describe dissipation in the corona and interaction of
the corona with the disc. 
Further work would be necessary to address 
relativistic particle acceleration and emission, 
illumination of the disc surface by X-rays produced in the corona
and subsequent heating of top layers of the disc, 
and emergence of magnetized outflows.

\section*{Acknowledgments}
V.I.~Shishov is thanked for the remark improving presentation of the results.
Two of authors (VP and EB) want to 
acknowledge their stay at the Institute of Theoretical Physics in Santa 
Barbara, where part of this work was done. SB acknowledges his stay at the 
University of Rochester when this work was initiated.
This research was supported in part by the National Science Foundation Under
Grant No. PHY99-07949.
VP and EB were also supported by DOE grant DE-FG02-00ER54600.

\appendix

\section[]{On cyclotron emission}

Since the characteristic temperature inside the disc ($T_{\rm {mpd}}$
given by equation~(\ref{eqn3.8})) is non-relativistic, cyclotron emission of an 
electron occurs at frequencies close to multiples of the gyrofrequency 
$\omega= s\omega_B$, where $s=1,2,3,\ldots$, and 
\begin{equation}
\omega_B=\frac{eB}{m_{\rm {e}} c}\approx 1.7\times 10^{11}\,\mbox{s}^{-1} 
\left(\frac{B_{10}}{10^4\,\mbox{G}}\right)\left(\frac{r}{10 r_{\rm {g}}}\right)^{-\delta}
\label{eqn4.1}\mbox{.}
\end{equation}
However, because the magnitude of the magnetic field varies across the disc, the 
resulting emission blends many discrete gyrolines. The typical
range of cyclotron emission is cm radio waves for AGNs and submillimetre to infrared 
for stellar mass black holes. Characteristic plasma parameters in our disc for the case 
of supermassive black holes are similar to those encountered in solar chromosphere,
where plasma effects are important for the generation and propagation of radio waves
\citep{zheleznyakov70}.  
One should expect that collective plasma effects will influence the 
cyclotron radiative process at such low frequencies. The plasma frequency is
\begin{eqnarray}
&& \omega_{\rm {L}}=\sqrt{\frac{4\pi e^2 n}{m_{\rm {e}}}} = 
8.8\times 10^{11}\,\mbox{s}^{-1}
\left(\frac{l_{\rm {E}}}{2\epsilon}\right)^{-1}\left(\frac{B_{10}}{10^4\,
\mbox{G}}\right)^3 \times\nonumber\\
&& {\mathcal{G}}^{-1} M_8 \left(\frac{r}{10 r_{\rm {g}}}\right)^{3-3\delta}
\label{eqn4.2}
\end{eqnarray}
and the ratio of cyclotron to plasma frequencies is
\begin{eqnarray}
&& \frac{\omega_B}{\omega_{\rm {L}}}=2\times 10^{-1} 
\left(\frac{r}{10 r_{\rm {g}}}\right)^{-3+2\delta}
{\mathcal{G}} \times\nonumber\\
&& \left(\frac{B_{10}}{10^4\,\mbox{G}}\right)^{-2}
\left(\frac{l_{\rm {E}}}{2\epsilon}\right) M_8^{-1}
\label{eqn4.3}\mbox{.}
\end{eqnarray}
We see that typically $\omega_B \sim \omega_{\rm {L}}$, so that cyclotron emitted radiation can
not propagate for some disc parameters. 
However, even if $\omega_B \gg \omega_{\rm {L}}$, the plasma affects 
cyclotron radiation. As summarized in \citet{zhelezniakov96}
collective effects suppress the emission on the first harmonic, $s=1$, such that it becomes
of order of the emission on the second harmonic, $s=2$. The emissivity on higher harmonics,
$s>2$, is smaller by the factor $(kT/m_{\rm {e}}c^2)^{s-2}$ 
than on the second and first harmonics.
This occurs only at high enough plasma densities $\displaystyle n \gg \frac{B^2}{4\pi}
\frac{3kT}{m_{\rm {e}}^2 c^4}$, which translates into $c^2 \gg v_{\rm {A}} c_{\rm {s}} 
m_{\rm {p}}/m_{\rm {e}}$, 
where $c_{\rm {s}}\sim 3kT/m_{\rm {p}}$. In vacuum, note that the emissivity
on the first harmonic is $(m_{\rm {e}} c^2/kT)$ times larger 
than the emissivity on second harmonic.
The latter condition is narrowly satisfied for small radii 
of our optically thick disc models and the larger the $r$ the better it is satisfied.

Cyclotron self-absorption also occurs in narrow lines centred on multiples of 
$\omega_B$. At some frequency $\omega$, emission and absorption occurs 
only in spatially narrow resonant layers inside the disc, 
where the magnetic field strength matches 
the frequency, i.e. $s\omega_B(B)-\omega$ is small. The width of emission and absorption 
frequency intervals is determined mainly by the thermal Doppler shifts 
$\displaystyle\Delta\omega/\omega \sim \sqrt{kT/m_{\rm {e}} c^2}$ \citep{zhelezniakov96}.
The width of such resonant layers can be estimated as $\sim H 
\sqrt{kT/m_{\rm {e}} c^2}$. 
\citet{zhelezniakov96} (chapter~6) gives the expression for the optical thickness through
such gyroresonance layers on the second harmonic $\omega=2\omega_B$, which
for our disc is 
\begin{eqnarray}
&& \tau_{\rm {cyc}}\approx \frac{\omega}{c}\frac{\omega_{\rm {L}}^2}{\omega^2}
\frac{kT}{m_{\rm {e}} c^2} H =
1.8 \times 10^{12} \left(\frac{B_{10}}{10^4\,\mbox{G}}\right)^4
\times \nonumber\\
&& \left(\frac{r}{10 r_{\rm {g}}}\right)^{39/8 -4\delta} {\mathcal{G}}^{-1} 
\left(\frac{l_{\rm {E}}}{2\epsilon}\right)^{-1} M_8^{9/4}
\label{eqn4.4}\mbox{.}
\end{eqnarray}
For all our models $\tau_{\rm {cyc}}$ is always 
very large and the emission is always strongly
self-absorbed. Cyclotron photons are not subject to Compton scattering by free electrons,
since the wavelength of the emission is always larger than Debeye radius in plasma, so
electrostatic shielding of charges prevents them from scattering. Under such circumstances
the resulting cyclotron flux from each gyroresonance layer is that of the black body 
with the local plasma temperature in the gyroresonance layer. Due to the overlapping 
of all layers, the resulting spectrum is a black body spectrum of 
width $\sim 2\omega_B$. Since $\hbar\omega_B \ll kT$, 
the total flux of cyclotron emission 
from the disc surface is negligibly small. 

\section[]{Star-disc collisions as possible source of turbulence}

When a star passes through a disc, it creates strong cylindrical shock 
propagating in the surrounding gas in the disc. The aftershock gas is heated to  
temperatures exceeding the equilibrium temperature in the accretion disc. As the 
shock weakens, this heating decreases until at some distance from the impact 
point the incremental heating becomes comparable to the equilibrium heat content.
The scale substantially affected by a star passage is 
$x\approx R_{\ast} v_{\rm {K}}/c_{\rm {s}}$, 
much larger than the radius of the star $R_{\ast}$.
The shock front can become unstable and turbulence can occur in the aftershock gas.
The heated gas becomes buoyant, rises above the disc and falls back because of gravity.
Fall-back occurs with supersonic velocities and can further excite turbulence. 
Turbulence will derive energy from both heating by star passages 
and shear of the flow. The energy, which can be derived from shear, is equal to $Q$ 
given by expression~(\ref{eqn11a}). It is possible that star-disc collisions might
mainly be a trigger for the available shear energy to be converted into supersonic 
turbulence, and additional energy deposited into the disc by star-disc collisions
is negligible. However, it seems unreasonable that
the star-disc collisions can influence the global structure of the accretion disc
unless the energy deposition from them is some fraction of the energy necessary to 
sustain turbulence level $Q$ in the disc. 

The energy deposition rate by stars per unit surface of the disc is
\begin{equation}
Q_{\ast}\approx \pi R_{\ast}^2 \Sigma v_{\ast}^2\frac{1}{2}n_{\ast}v_{\ast}
\label{eqn5.1}\mbox{,}
\end{equation}
where $v_{\ast}\approx v_{\rm {K}}$ is the typical velocity of stars at radial distance $r$,
$R_{\ast}\approx R_{\sun}$ is the average radius of stars, $n_{\ast}=n_{\ast}(r)$ 
is the number density of stars, $\Sigma$ is the surface density of the disc given 
by expression~(\ref{eqn18}) in our model. Only accretion discs orbiting 
supermassive black holes $M \ga 10^6\, {\rm M}_{\sun}$ can be 
influenced by star-disc collisions.

The resolution of observations is only enough to estimate the
number density of stars at about 1~pc for M32 and M31 and about 10~pc for 
nearest ellipticals. In line with these observations we assume a star density 
$n(\mbox{1 pc})\approx 10^4 - 10^6\, {\rm M}_{\sun} \mbox{pc}^{-3}$ at 1 pc distance 
from the central massive black hole \citep{lauer95}.
To estimate $n_{\ast}$ for $r\leq 10^3 r_{\rm {g}}$ 
we need to rely on the theory of central 
star cluster evolution. 
The gravitational potential inside the central 1~pc will be always dominated 
by the black hole. \citet{bahcall76} showed that, if the evolution of 
a star cluster is dominated by relaxation,
the effect of a central Newtonian point mass on an isotropic cluster would
be to create a density profile $n \propto r^{-7/4}$. However, for small
radii ($\approx 0.1-1\,\mbox{pc}$) the physical collisions of stars
dominate two-body relaxations. Also, regions 
near the black hole will be devoid of stars due to tidal disruption and 
capture by the black hole. Numerical simulations 
of the evolution of the central star cluster, taking into account 
star-star collisions, star-star gravitational interactions,
tidal disruptions and relativistic effects were recently 
performed by \citet{rauch99}. \citet{rauch99} showed that star-star
collisions lead to the formation of a plateau in stellar density for small $r$
because of the large rates of destruction by collisions.
We adopt the results of model~4 from \cite{rauch99} as our fiducial model.
This model was calculated for all stars having initially one solar mass.
The collisional evolution is close to a stationary state,
when the combined losses of stars due to collisions, ejection, tidal
disruptions and capture by the black hole are balanced by the replenishment
of stars as a result of two-body relaxation in the outer region 
with a $n\propto r^{-7/4}$ density profile. Taking into account the 
order of magnitude uncertainty in the the observed star density at 1~pc, 
the fact that model~4 has not quite reached 
a stationary state can be accepted for order of magnitude estimates.
For $M=10^8 M_8 \,{\rm M}_{\sun}$ we approximate the 
density profile of model~4 as 
\begin{eqnarray}
&& n=n_5 \times 10^5 \, \frac{{\rm M}_{\sun}}{\mbox{pc}^3}\, \left(\frac{r}{1\mbox{pc}}
\right)^{-7/4} \quad \mbox{for} \quad r>10^{-2}\,\mbox{pc}\mbox{,} 
\nonumber \\
&& n=n_5 \times 3\times 10^8\,\frac{{\rm M}_{\sun}}{\mbox{pc}^3} \quad \mbox{for}
\quad     5 r_{\rm {t}}<r<10^{-2}\,\mbox{pc}\label{eqn5.2}\mbox{,} \\
&& n \quad \mbox{decreases} \quad \mbox{for} \quad r< 5 r_{\rm {t}}\mbox{,} \nonumber
\end{eqnarray}
where $r_{\rm {t}}=2.1\times 10^{-4}\,\mbox{pc}\times M_8^{1/3} = 
20 r_{\rm {g}} M_8^{-2/3}$ 
is the tidal disruption radius for a solar mass star, 
and $\displaystyle n_5=\frac{n(1\,\mbox{pc})}{
10^5\,{\rm M}_{\sun}/\mbox{pc}^{-3}}$. 
The region $r<r_{\rm {t}}$ is completely devoid of stars.

We see that star-disc collisions cannot excite turbulence and strong magnetic 
fields in the very inner part of the accretion disc, for $r<r_{\rm {t}}$, and such excitation
should be weak for $r_{\rm {t}} < r < 5 r_{\rm {t}}$. 
The relative width of the star depleted region,
$5 r_{\rm {t}}/r_{\rm {g}}$, decreases with increasing $M$. 
For $M\approx 2\times 10^9\, {\rm M}_{\sun}$
$r_{\rm {t}} = 3 r_{\rm {g}}$ and star-disc collisions happen all over the disc. For 
$M < 3\times 10^6\, {\rm M}_{\sun}$ $5 r_{\rm {t}} > 10^3 r_{\rm {g}}$ 
and for $M < 3\times 10^5\, {\rm M}_{\sun}$
$r_{\rm {t}} > 10^3 r_{\rm {g}}$ and star-disc collisions are unimportant for the structure of 
the accretion disc. Let us assume that it should be $Q_{\ast}=f Q$ where the fraction 
$f$ is less than unity but not much less than unity. Further, we use 
expression~(\ref{eqn11a}) for $Q$, expression~(\ref{eqn18}) for $\Sigma$, and
the value for $n$ in the constant density core of the star cluster, second raw 
in expression~(\ref{eqn5.2}), to substitute into $Q_{\ast}=f Q$. 
Since the relation $Q_{\ast}=f Q$ should be satisfied
for all values of $r$, the value of $\delta$ is determined and turns out to be 
$\delta=3/2$. Solving the rest of the equation for the magnitude of magnetic field 
$B_{10}$ at $\delta=3/2$ we obtain
\begin{equation}
\frac{B_{10}}{10^4\,\mbox{G}}=6.5\times 10^4\, 
\left(\frac{l_{\rm {E}}}{2\epsilon}\right)^{1/2}
M_8^{-3/4} n_5^{-1/4} f^{1/4}\label{eqn5.3}\mbox{.}
\end{equation}
This value of $B$ required by energy input from star-disc collisions should fall into
the range of constraints for the $B_{10}$ listed in the end of section~\ref{sec3}.

We explored all feasible range of parameters $M_8$, $l_{\rm {E}}/\epsilon$, $f>10^{-3}$,
$n_5 < 10^3$ and found that the magnetic field calculated from 
expression~(\ref{eqn5.3}) is always too strong to fall in the allowed range of parameters 
discussed at the end of section~\ref{sec3}. In particular, the constraint that 
magnetic and turbulent pressure dominate thermal and radiation pressure is
violated. The minimum number density of stars necessary to satisfy this constraint 
at the most favourable values of other parameters still plausible for some AGN 
($M_8 =40$, $l_{\rm {E}}/\epsilon = 10^{-10}$, $f=10^{-3}$), turns out to be 
$n_5 \approx 10^6$. Such a high number density of stars would imply total mass
in stars of order of $10^{11} {\rm M}_{\sun}$ inside the central parsec from the 
central black hole. This mass exceeds observational and theoretical 
limitations.


\begin{thebibliography}{}

\bibitem[\protect\citeauthoryear{Armitage, Reynolds \& Chiang}{Armitage et al.}{2001}]%
{armitage01} Armitage~P.~J., 
Reynolds~C.~S., Chiang~J., 2001, ApJ, 548, 868

\bibitem[\protect\citeauthoryear{Bahcall \& Wolf}{1976}]{bahcall76} 
Bahcall~J.~N., Wolf~R.~A., 1976, ApJ, 209, 214

\bibitem[\protect\citeauthoryear{Balbus \& Hawley}{1998}]{balbus98} 
Balbus~S.~A., Hawley~J.~F., 1998, Rev. of Modern Physics, 70, 1

\bibitem[\protect\citeauthoryear{Beloborodov}{1999}]{beloborodov99}
Beloborodov~A.M., 1999, in Poutanen~J., Svensson~R., eds, ASP Conf.
Ser. Vol.~161, High Energy Processes in Accreting Black Holes.
Astron. Soc. Pac., San Francisco, p.~295

\bibitem[\protect\citeauthoryear{Blackman}{2002}]{blackman02}
Blackman~E.~G., 2002, to appear in the proceedings of the 1st Niels Bohr Summer 
Institute: "Beaming and Jets in Gamma-Ray Bursts", Copenhagen, Aug.
2002 (astro-ph/0211187)

\bibitem[\protect\citeauthoryear{Blackman \& Pariev}{2003}]{blackman03}
Blackman~E.~G., Pariev~V.~I., 2003, in preparation

\bibitem[\protect\citeauthoryear{Brandenburg}{1998}]{brandenburg98} Brandenburg~A.,
1998, in Abramowicz~M.~A., Bjornsson~G., Pringle~J.~E., eds, 
Theory of Black Hole Accretion Discs.
Cambridge University Press, Cambridge, p.~61

\bibitem[\protect\citeauthoryear{Brandenburg et al.}{1995}]{brandenburg95}
Brandenburg~A., Nordlund~A., Stein~R.F., Torkelsson~U., 1995, 
ApJ, 446, 741

\bibitem[\protect\citeauthoryear{Campbell}{2000}]{campbell00} Campbell~C.~G., 
2000, MNRAS, 317, 501

\bibitem[\protect\citeauthoryear{Di~Matteo, Blackman \& Fabian}{Di~Matteo et al.}{1997a}]%
{dimatteo97a} Di Matteo~T., Blackman~E.G., Fabian~A.~C., 1997a, MNRAS, 
291, L23

\bibitem[\protect\citeauthoryear{Di~Matteo, Celotti \& Fabian}
{Di~Matteo et al.}{1999}]{dimatteo99}  Di Matteo~T., Celotti~A., 
Fabian~A.~C., 1999, MNRAS, 304, 809

\bibitem[\protect\citeauthoryear{Done}{2002}]{done02} Done~C., 
2002, Roy. Soc. of London Phil. Tr.~A, 360, 1967

\bibitem[\protect\citeauthoryear{Eardley \& Lightman}{1975}]{eardley75} 
Eardley~D.~M., Lightman~A.~P., 1975, ApJ, 200, 187

\bibitem[\protect\citeauthoryear{Field \& Rogers}{1993a}]{field93a} 
Field~G.~B., Rogers,~R.~D., 1993a, ApJ, 403, 94

\bibitem[\protect\citeauthoryear{Field \& Rogers}{1993b}]{field93b}  
Field~G.~B., Rogers~R.~D., 1993b, unpublished

\bibitem[\protect\citeauthoryear{Gammie}{2000}]{gammie00} 
Gammie~C.~F., 2000, ApJ, 522, L57

\bibitem[\protect\citeauthoryear{Ikhsanov}{1989}]{ikhsanov89} 
Ikhsanov~N.~R., 1989, Soviet Astr. Lett., 15, 220

\bibitem[\protect\citeauthoryear{Krolik}{1999}]{krolik99} 
Krolik~J.~H., 1999, Active Galactic Nuclei: From the 
Central Black Hole to the Galactic Environment. 
Princeton University Press, Princeton

\bibitem[\protect\citeauthoryear{Hawley}{2001}]{hawley01b} Hawley~J.~F.,
2001, ApJ, 554, 534

\bibitem[\protect\citeauthoryear{Hawley \& Krolik}{2001}]{hawley01a} 
Hawley~J.~F., Krolik~J.~H., 2001, ApJ, 548, 348

\bibitem[\protect\citeauthoryear{Lauer et al.}{1995}]{lauer95} 
Lauer~T.~R., Ajhar~E.~A., Byun~Y.-I., Dressler~A.,
Faber~S.~M., Grillmair~C., Kormendy~J., Richstone~D., Tremaine~S.,
1995, AJ, 110, 2622

\bibitem[\protect\citeauthoryear{Lynden-Bell}{1969}]{lyndenbell69} 
Lynden-Bell~D., 1969, Nature, 223, 690

\bibitem[\protect\citeauthoryear{Machida, Hayashi \& Matsumoto}{Machida et al.}%
{2000}]{machida00} 
Machida~M., Hayashi~M.~R., Matsumoto~R., 2000, ApJ, 532, L67

\bibitem[\protect\citeauthoryear{Merloni}{2003}]{merloni03}
Merloni~A., 2003, MNRAS, 341, 1051

\bibitem[\protect\citeauthoryear{Merloni \& Fabian}{2001}]{merloni01}
Merloni~A., Fabian~A.C., 2001, MNRAS, 321, 549

\bibitem[\protect\citeauthoryear{Miller \& Stone}{2000}]{miller00} 
Miller~K.~A., Stone~J.~M., 2000, ApJ, 534, 398

\bibitem[\protect\citeauthoryear{Mushotzky, Done \& Pounds}{Mushotzky et al.}%
{1993}]{mushotzky93} Mushotzky~R.F., Done~C., Pounds~K.A., 1993,
Annu. Rev. Astron. Astrophys., 31, 717

\bibitem[\protect\citeauthoryear{Nowak}{1995}]{nowak95} Nowak~M.A.,
1995, PASP, 107, 1207

\bibitem[\protect\citeauthoryear{Ogilvie \& Livio}{2001}]{ogilvie01} 
Ogilvie~G.~I., Livio~M., 2001, ApJ, 553, 158

\bibitem[\protect\citeauthoryear{Ostriker, Stone \& Gammie}{Ostriker et al.}%
{2001}]{ostriker01} Ostriker~E.~C., 
Stone~J.~M., Gammie~C.F., 2001, ApJ, 546, 980

\bibitem[\protect\citeauthoryear{Rauch}{1999}]{rauch99} Rauch~K.~P., 
1999, ApJ, 514, 725

\bibitem[\protect\citeauthoryear{Rybicki \& Lightman}{1979}]{rybicki79} 
Rybicki~G.~B., Lightman~A.~P.,
1979, Radiative Processes in Astrophysics. John Wiley, New York

\bibitem[\protect\citeauthoryear{Shakura}{1972}]{shakura72} Shakura~N.~I.,
1972, AZh, 49, 921

\bibitem[\protect\citeauthoryear{Shakura \& Sunyaev}{1973}]{shakura73} 
Shakura~N.~I., Sunyaev~R.~A., 1973, A\&A, 24, 337

\bibitem[\protect\citeauthoryear{Shalybkov \& R\"udiger}{2000}]{shalybkov00} 
Shalybkov~D., R\"udiger~G., 2000, MNRAS, 315, 762

\bibitem[\protect\citeauthoryear{Shapiro \& Teukolsky}{1983}]{shapiro83} 
Shapiro~S.~L., Teukolsky~S.~A., 1983, Black Holes, White Dwarfs, and Neutron Stars.
John Wiley, New York

\bibitem[\protect\citeauthoryear{Shapiro, Lightman \& Eardley}{Shapiro et al.}%
{1976}]{shapiro76} Shapiro~S.L., Lightman~A.P., Eardley~D.M., 1976, ApJ,
204, 187 

\bibitem[\protect\citeauthoryear{Shibata, Tajima \& Matsumoto}{Shibata et al.}%
{1990}]{shibata90} 
Shibata~K., Tajima~T., Matsumoto~R., 1990, ApJ, 350, 295

\bibitem[\protect\citeauthoryear{Stone}{1999}]{stone99} 
Stone~J.~M., 1999, in Franco~J., Carrami\~nana~A., eds, 
Interstellar Turbulence.
Cambridge University Press, Cambridge, p.~267

\bibitem[\protect\citeauthoryear{Stone, Ostriker \& Gammie}{Stone et al.}%
{1998}]{stone98} Stone~J.~M., 
Ostriker~E.~C., Gammie~C.~F., 1998, ApJ, 508, L99

\bibitem[\protect\citeauthoryear{Wandel \& Petrosian}{1988}]{wandel88} 
Wandel~A., Petrosian~V., 1988, ApJ, 329, L11

\bibitem[\protect\citeauthoryear{Zhelezniakov}{1996}]{zhelezniakov96} 
Zhelezniakov~V.~V., 1996, Radiation in astrophysical plasmas. 
Kluwer, Dordrecht 

\bibitem[\protect\citeauthoryear{Zheleznyakov}{1970}]{zheleznyakov70} 
Zheleznyakov~V.~V., 1970, Radio Emission of the Sun and Planets. 
Pergamon Press, Oxford

\end{thebibliography}
\end{document}